\documentclass[aps,twocolumn,prd,preprintnumbers,superscriptaddress,nofootinbib,bibnotes,floatfix,longbibliography]{revtex4-2}

\pdfoutput=1
\usepackage{amsmath}
\usepackage{amsfonts}
\usepackage{amssymb}
\usepackage{mathrsfs}
\usepackage{graphicx}
\usepackage{color}
\usepackage{bbding}
\usepackage{tabularx}
\usepackage[usenames,dvipsnames,table]{xcolor}
\usepackage[normalem]{ulem}
\usepackage{makecell}
\usepackage{siunitx}
\usepackage{enumerate}
\usepackage{multirow}
\usepackage{makecell}
\usepackage{upgreek}
\usepackage{comment}
\usepackage{subfigure}

\usepackage{marvosym}
\definecolor{darkred}{rgb}{0.5,0,0}
\definecolor{darkblue}{rgb}{0,0,0.5}
\definecolor{firebrick}{rgb}{0.75,0.125,0.125}
\definecolor{darkgreen}{rgb}{0,0.5,0}
\usepackage[colorlinks=true,linkcolor=firebrick,citecolor=darkgreen,urlcolor=darkblue]{hyperref}

\newcommand{\beq}{\begin{equation}}
\newcommand{\eeq}{\end{equation}}
\newcommand{\ie}{{i.e.}}

\newcommand{\eg}{{e.g.}}

\newcommand{\eq}{Eq.}

\newcommand{\fig}{Fig.}

\newcommand{\Refe}{Ref.}
\newcommand{\Refes}{Refs.}

\newcommand{\equ}[1]{\eq~(\ref{equ:#1})}
\newcommand{\figu}[1]{\fig~\ref{fig:#1}}

\graphicspath{{figures/}}

\newcommand{\orcid}[1]{\href{https://orcid.org/#1}{\includegraphics[width=10pt]{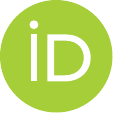}}}

\begin{document}

\title{Probing Lorentz invariance with a high-energy neutrino flare}

\author{Mauricio Bustamante \orcid{0000-0001-6923-0865}}
\email{mbustamante@nbi.ku.dk}
\affiliation{Niels Bohr International Academy, Niels Bohr Institute,\\University of Copenhagen, DK-2100 Copenhagen, Denmark}

\author{John Ellis \orcid{0000-0002-7399-0813}}
\email{john.ellis@cern.ch}
\affiliation{Theoretical Physics and Cosmology Group, Department of Physics, King’s College London, London WC2R 2LS, UK}
\affiliation{Theoretical Physics Department, CERN, CH-1211 Geneva 23, Switzerland}

\author{Rostislav Konoplich \orcid{0000-0002-6223-7017}}
\email{rostislav.konoplich@manhattan.edu}
\affiliation{Department of Mathematics and Physics, Manhattan University,\\
4513 Manhattan College Parkway, Riverdale, NY 10471, United States of America}
\affiliation{Department of Physics, New York University,\\
726 Broadway, New York, NY 10003, United States of America}

\author{Alexander S.~Sakharov \orcid{0000-0001-6622-2923}}
\email{alexandre.sakharov@cern.ch}
\affiliation{Department of Mathematics and Physics, Manhattan University,\\
4513 Manhattan College Parkway, Riverdale, NY 10471, United States of America}
\affiliation{Experimental Physics Department, CERN, CH-1211 Gen\`eve 23, Switzerland}

\date{August 28, 2024}
%\date{today}

\begin{abstract}

Time-of-flight measurements of high-energy astrophysical neutrinos can be used to probe
Lorentz invariance, a pillar of modern physics. If Lorentz-invariance violation (LIV) occurs, it could cause neutrinos to slow down, with the delay scaling linearly or quadratically with their energy.
We introduce non-parametric statistical methods designed to detect LIV-induced distortions in the temporal structure of a high-energy neutrino flare as it travels to Earth from a distant astrophysical source, independently of the intrinsic timing properties of the source.
Our approach, illustrated using the 2014/2015 TeV–PeV neutrino flare from the blazar TXS 0506+056 detected by IceCube, finds that the LIV energy scale must exceed $10^{14}$~GeV (linear) or $10^9$~GeV (quadratic). Our methods provide a robust means to investigate LIV by focusing solely on a neutrino flare, without relying on electromagnetic counterparts, and account for realistic energy and directional uncertainties. For completeness, we compare our limits inferred from TXS 0506+056 to the sensitivity inferred from multi-messenger detection of tentative coincidences between neutrinos and electromagnetic emission from active galactic nuclei and tidal disruption events.
\\
~~\\
KCL-PH-TH/2024-22, CERN-TH-2024-056

\end{abstract}

\maketitle

%%%%%%%%%%%%%%%%%%%%%%%%%%%%%%%%%%%%%%%%%%%%%%%%%%%%%%%%%%%%%%%%%%%%%%%%%%%%
%%%%%%%%%%%%%%%%%%%%%%%%%%%%%%%%%%%%%%%%%%%%%%%%%%%%%%%%%%%%%%%%%%%%%%%%%%%%

\section{Introduction}
\label{sec:intro}

Lorentz invariance is one of the most fundamental principles
of physics, underlying the formulations of special and
general relativity. It therefore behooves physicists to test
its validity as accurately as possible.
Over the past quarter-century there has been widespread
interest~\cite{Coleman:1997xq,Amelino-Camelia:1997ieq,Colladay:1998fq,Coleman:1998ti}
in the possibility that it might be violated
in the propagation of particles at high energies.  Their velocities, $v$, might
differ from that of light, $c$, by some energy-dependent amount, \ie,
$\Delta v \propto (E/M_{p,n})^n$,
where $E$ is the energy of a particle of type $p$, and $n$ and $M_{p,n}$ are model-dependent parameters; $n = 1$ or 2 are commonly considered.
General theoretical~\cite{Amelino-Camelia:1997ieq}
and phenomenological arguments~\cite{Cohen:2011hx} suggest that particle velocities
might be reduced at higher energies, but not increased.

%%%%%%%%%%%%%%%%%%%%%%%%%%%%%%%%%%%%%%%%%%%%%%%%%%%%%%%%%%%%%%%%%%%%%%%%%%%%
\begin{figure}[t!]
 \centering
 \includegraphics[width=1.0\columnwidth]{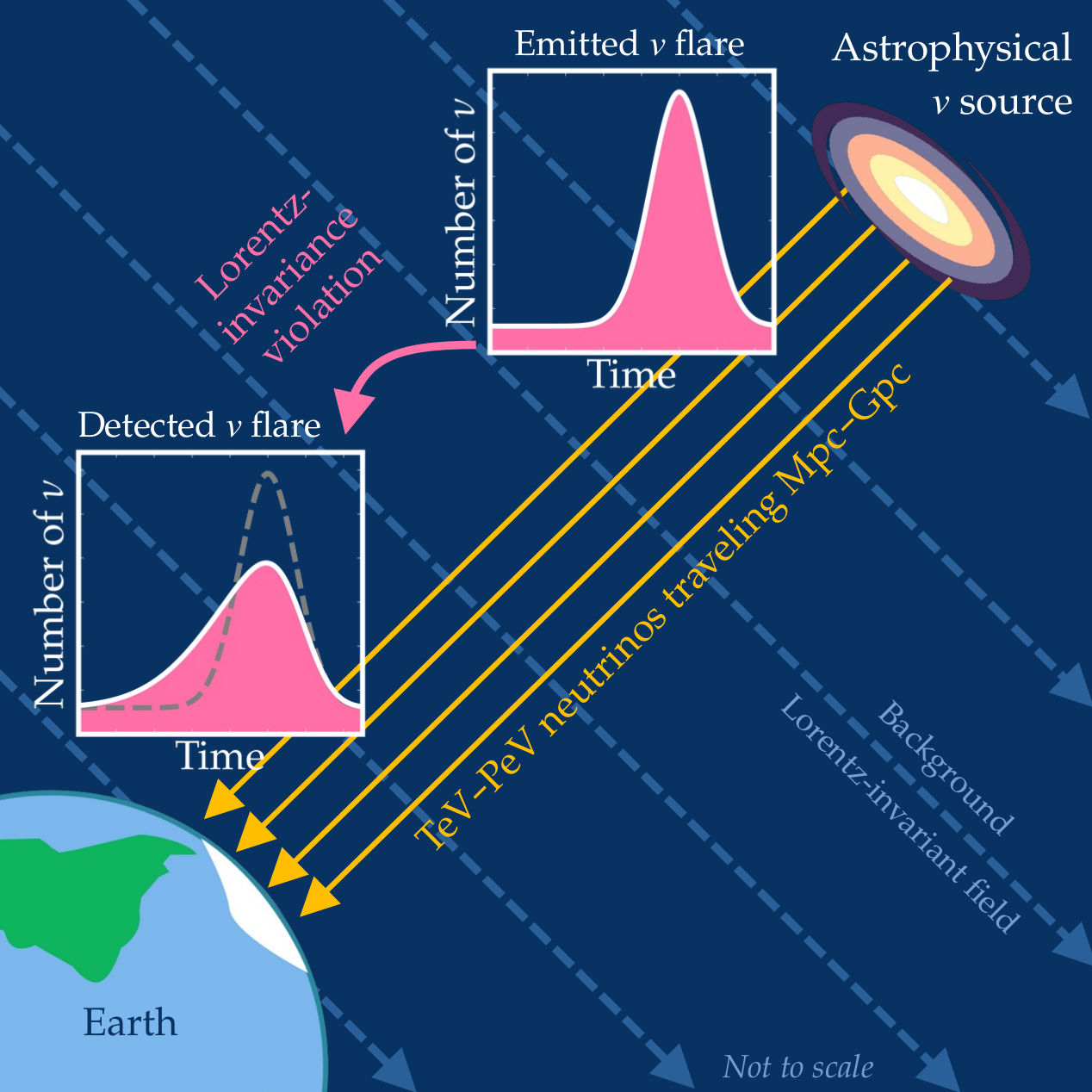}
 \caption{\label{fig:overview}
 \textbf{\textit{Broken Lorentz invariance changes the time distribution of a flare of high-energy astrophysical neutrinos.}}  Higher-energy neutrinos emitted by a transient source, like the blazar TXS 0506+056, would travel more slowly than lower-energy neutrinos.  Their time distribution grows more uniform, flatter, and more skewed.  \textbf{\textit{The absence of these features in a detected flare places robust limits on the energy scale of Lorentz-invariance violation.}}
 \vspace*{-1.2cm}
 }
\end{figure}
%%%%%%%%%%%%%%%%%%%%%%%%%%%%%%%%%%%%%%%%%%%%%%%%%%%%%%%%%%%%%%%%%%%%%%%%%%%%

Reference~\cite{Amelino-Camelia:1997ieq} suggested that
the most stringent tests of this form of Lorentz-invariance violation (LIV) would be provided by
transient astrophysical phenomena that emit high-energy photons, and possibly also protons and neutrinos, with TeV--PeV energies and above, such as pulsars, gamma-ray bursts (GRBs) and flaring
active galactic nuclei (AGNs).
The effects of LIV would accumulate over the long distances traveled by
the particles from their astrophysical sources to Earth, becoming detectable.
Lorentz-invariance violation would manifest itself in differences in the
arrival times of particles of different energies or types
emitted at similar times.

Observations
of high-energy gamma rays from GRBs and AGNs have provided the most stringent and robust
probes of Lorentz invariance in photon propagation (see \Refe~\cite{Piran:2023xfg} and references therein), while
multi-messenger observations of a binary neutron star merger
have provided a first direct constraint on the possible difference
in velocities of electromagnetic and gravitational waves~\cite{LIGOScientific:2017zic}, and observations of gravitational waves over a range of frequencies
have constrained the possibility of LIV in their
propagation~\cite{Ellis:2016rrr, LIGOScientific:2019fpa}.

Because neutrinos are notoriously difficult to detect, testing Lorentz invariance in their propagation is challenging.
Nevertheless, the
observation of neutrinos from supernova SN 1987A
constrained a possible energy dependence of
neutrino velocities~\cite{Ellis:2008fc}, accelerator neutrinos explored alternative probes of LIV~\cite{MINOS:2012ozn, MINOS:2015iks}, and multi-messenger observations of the
blazar TXS~0506+056 have been used to constrain a possible
difference between the velocities of photon and neutrino
propagation~\cite{Ellis:2018ogq}. However, the latter constraint relies upon the detection by the
IceCube neutrino telescope in 2017 of a single high-energy neutrino in coincidence with a gamma-ray flare observed by several
electromagnetic detectors~\cite{IceCube:2018dnn}.

Subsequent inspection of archival IceCube data revealed a flare of multiple TeV-scale neutrinos in 2014/2015 from the direction of TXS 0506+056~\cite{IceCube:2018cha}, lending further credence to its being the first high-energy neutrino source discovered. (Yet, the original pre-trial significance of this detection, of $7.0 \cdot 10^{-5}$~\cite{IceCube:2018cha},
was reduced to $8.1 \cdot 10^{-3}$~\cite{IceCube:2021xar} after a re-evaluation of the data~\cite{IceCube:2021xar, IceCube:2023oua}.)
The 2014/2015 neutrino flare was seemingly not accompanied by a gamma-ray flare (see, however, \Refe~\cite{Fermi-LAT:2019hte}), unlike the 2017 flare, complicating the modeling of TXS 0506+056 as a multi-messenger source; see, \eg, \Refes~\cite{Cerruti:2018tmc, Keivani:2018rnh, Murase:2018iyl, Halzen:2018iak, Reimer:2018vvw, Rodrigues:2018tku, Padovani:2019xcv, Xue:2019txw}.
However, our analysis is decoupled from such complications, since it relies exclusively on the observation of the flare of high-energy neutrinos.

Figure~\ref{fig:overview} illustrates how the detection of a high-energy neutrino flare from a transient astrophysical source can facilitate new tests of Lorentz invariance. As neutrinos travel to Earth, their energy-dependent slow-down alters their arrival times, reshaping the time distribution of the neutrino in the flare into
one that is flatter, less peaked, and more skewed than the original emission.  This transformation occurs regardless of the initial time distribution, thereby  making our analysis immune to uncertainties in modeling neutrino production.

We look for these features using techniques originally developed for gamma rays from GRBs~\cite{Ellis:2018lca}.   \textbf{\textit{We show how to use a high-energy neutrino flare to constrain Lorentz-invariance violation in neutrino propagation even in the absence of a counterpart electromagnetic flare}}. As illustration, we use the TXS 0506+056 flare to place new limits on the energy scale of LIV that is linearly ($n = 1$) or quadratically ($n = 2$) dependent on neutrino energy.

Our methods avoid the substantial uncertainties intrinsic to relying on a single high-energy neutrino to test Lorentz invariance~\cite{Ellis:2018ogq}.  However, unlike studies based on gamma rays, where the energy resolution is excellent, the sensitivity of our analysis is hampered by the challenge in reconstructing the energy of detected high-energy astrophysical neutrinos, which we account for.  Nonetheless, our results are strengthened by accounting for realistic experimental details.

This paper is structured as follows. Section~\ref{sec:LVmotivate} reviews the motivation for studying the time of flight of high-energy radiation
from astrophysical sources to search for LIV.
Section~\ref{sec:flares} describes the high-energy neutrino data detected by IceCube during the 2014/2015 TXS 0506+056 flare.
Section~\ref{sec:methods} describes and adapts methods previously developed for studying GRBs~\cite{Ellis:2018lca} to analyze neutrino flares.
Section~\ref{sec:liv_constraints_txs} evaluates the sensitivity of the neutrino flare to LIV.
Section~\ref{sec:other_limits} compares this sensitivity with the other sensitivities outlined above.
Section~\ref{sec:summary} summarizes and concludes.

%%%%%%%%%%%%%%%%%%%%%%%%%%%%%%%%%%%%%%%%%%%%%%%%%%%%%%%%%%%%%%%%%%%%%%%%%%%%
%%%%%%%%%%%%%%%%%%%%%%%%%%%%%%%%%%%%%%%%%%%%%%%%%%%%%%%%%%%%%%%%%%%%%%%%%%%%

\section{Lorentz-invariance violation in high-energy astrophysical neutrinos}
\label{sec:LVmotivate}

%%%%%%%%%%%%%%%%%%%%%%%%%%%%%%%%%%%%%%%%%%%%%%%%%%%%%%%%%%%%%%%%%%%%%%%%%%%%

\subsection{Overview}
\label{sec:LVmotivate-overview}

Lorentz invariance underlies the very
successful geometric picture of space-time that special and general relativity embody. Although it has been verified
to high accuracy in many laboratory measurements~\cite{Kostelecky:2008ts}, there is no guarantee that it persists
at very short distances or to very high energies.
Indeed, the possibility of LIV at high
energies has been raised from several points of view~\cite{Coleman:1997xq, Amelino-Camelia:1997ieq, Colladay:1998fq, Coleman:1998ti}.

The motivation for this possibility proposed
in~\Refe~\cite{Amelino-Camelia:1997ieq} came from the idea that
space-time should be regarded as a dynamical medium with
a foamy short-distance structure generated by quantum-gravitational
fluctuations that appear and disappear on a Planckian time-scale~\cite{Wheeler:1957mu}.
This idea of \textit{space-time foam} has been developed heuristically
but not derived mathematically from first principles; see \Refe~\cite{Carlip:2022pyh} for a review.

According to the intuition underlying this approach, a passing
energetic particle interacts with this dynamical medium, distorting
the foamy space-time background. The back-reaction from the foam
modifies the propagation velocity of the particle, much as the
velocity of a photon is modified when it passes through a transparent
medium with an index of refraction induced by the interactions of the photon with
molecules in the medium. This intuition suggests that the dispersion
relation of the particle should be modified by its interaction with
the space-time foam by an amount $\propto (E/M_{\rm Pl})^n$, where $M_{\rm Pl} \sim 10^{18}$~GeV
is the reduced Planck mass.  The possibility
that $n = 1$ has been supported by heuristic estimates
inspired by string theory~\cite{Ellis:2008gg}.

The Standard Model Extension (SME)~\cite{Colladay:1998fq} takes a different approach
by extending the formulation of
conventional quantum field theory to include LIV
induced by spontaneous symmetry breaking in a Lorentz-invariant
fundamental theory. The SME has many desirable properties such as
gauge invariance, Hermiticity, causality, and power-counting
renormalizability, and may be considered independently from any
specific theoretical motivation. In this case, the violation of
Lorentz invariance in particles of type $p$ is $\propto (E/M_{p,n})^n$, where $n \ge 2$ is determined
by the choice of LIV operator.

Reference~\cite{Coleman:1997xq} takes yet another approach, drawing
attention to the possibility of LIV in neutrino oscillations.\footnote{Reference~\cite{Coleman:1998ti} applied the
same approach to the possible violation of the
Greisen-Zatsepin-Kuzmin cut-off on cosmic-ray energies~\cite{Greisen:1966jv, Zatsepin:1966jv}.}
This possibility, particularly in high-energy astrophysical neutrinos, has been studied extensively, \eg, in \Refes~\cite{Kostelecky:2003xn, Kostelecky:2003cr, Barenboim:2003jm, Kostelecky:2004hg, Katori:2006mz, Diaz:2009qk, Ando:2009ts, Bustamante:2010nq, Bhattacharya:2010xj, Katori:2011zz, Diaz:2011ia, Arguelles:2015dca, Bustamante:2015waa, Shoemaker:2015qul, IceCube:2017qyp, IceCube:2021tdn, Telalovic:2023tcb}.
Because transient high-energy neutrino emission is currently detected via a single flavor~\cite{IceCube:2018cha}, $\nu_\mu$ (Sec.~\ref{sec:flares}), we do not consider the effects of LIV in neutrino oscillations, but
assume for simplicity that it affects neutrinos of all flavors equally.

High-energy astrophysical neutrinos are particularly interesting probes of
Lorentz invariance.
Because, unlike energetic photons, neutrinos do not interact with cosmological photon backgrounds~\cite{Kifune:1999ex}, they reach us with energies beyond 100~TeV, where evidence for LIV might become prominent.
This long-recognized potential has been explored in \eg, \Refes~\cite{Barenboim:2003jm, Ando:2009ts, Bustamante:2010nq, Bhattacharya:2010xj, Arguelles:2015dca, Bustamante:2015waa, Shoemaker:2015qul, IceCube:2021tdn, Telalovic:2023tcb}.

%%%%%%%%%%%%%%%%%%%%%%%%%%%%%%%%%%%%%%%%%%%%%%%%%%%%%%%%%%%%%%%%%%%%%%%%%%%%

\subsection{Why high-energy astrophysical neutrino flares?}
\label{sec:LVmotivate-why_flares}

We focus on LIV that acts during the propagation of high-energy astrophysical neutrinos from their sources to Earth, across Mpc--Gpc distances, slowing them down and affecting their arrival times~\cite{Ellis:1999sf}.  In so doing, we follow the
suggestion for photons made in \Refe~\cite{Amelino-Camelia:1997ieq}, but avoid subscribing to a specific theoretical approach.

Traditionally, studies along these lines (Sec.~\ref{sec:other_limits-nu_gamma_associations}) have been based on
comparing the arrival times of a single high-energy neutrino---or handful of them---attributed to an alleged
astrophysical neutrino source \textit{vs.}~the arrival times of coincident electromagnetic emission from the same source~\cite{AlvesBatista:2023wqm}.
% The fundamental assumption in these studies is that neutrinos and photons
% are emitted by the same source at roughly the same time.
These studies are based on the assumption that neutrinos and photons are emitted by the source
simultaneously or nearly so. However, this assumption is difficult to justify, as various
astrophysical models predict intrinsic time offsets between neutrino and electromagnetic emissions.
Below, we outline some of these scenarios.

In choked jet models of
GRBs~\cite{Murase:2013ffa, Senno:2015tsn, Tamborra:2015fzv}, a jet forms but fails to
break through the stellar envelope, leading to neutrino production without a prompt electromagnetic signal.
Neutrinos are generated as accelerated particles within the choked jet interact with ambient matter, potentially
resulting in an early neutrino signal that precedes any electromagnetic emission from other processes in the system.
A key prediction of this model is that the majority of neutrinos will act as precursors to
the prompt gamma-ray emission. Therefore, in a neutrino electromagnetic coincidence search,
it is essential to analyze time windows about $10^3$ seconds before the gamma rays trigger~\cite{Senno:2015tsn}.
In blazars, high-energy neutrinos can be produced in photohadronic interactions between accelerated protons and
ambient photons, potentially in different emission zones. This can lead to time delays between neutrino and
gamma-ray flares~\cite{Gao:2016uld,Keivani:2018rnh}, depending on particle travel times within the jet and
other regions of the AGN. In the case of tidal disruption events (TDEs), where a star is torn apart by
a supermassive black hole, neutrino production may occur later in the disruption process,
following the initial electromagnetic outburst. For instance, neutrinos can be generated when debris from the
disrupted star falls back toward the black hole, interacting with accretion flows or jets that
form later in the event~\cite{Stein:2020xhk}. In two-zone emission models for GRBs and blazars,
where different regions within the same source are responsible for producing neutrinos and electromagnetic radiation,
inherent delays can arise. Neutrinos may be produced in a region with denser target material, while
gamma-rays emerge from a separate, often less-dense zone, leading to potential time offsets~\cite{Gao:2018mnu}.

The above examples highlight the complexity of neutrino - electromagnetic messengers associations in their relative temporal production and underscore the importance of considering timing offsets in multi-messenger
studies of LIV. In fact, the sensitivity of these studies, when performed properly, is significantly weakened by the challenge of disentangling time delays that are intrinsic to the source from those that could be induced by LIV during propagation~\cite{Levy:2024eiq}. Further complications are detailed in Sec.~\ref{sec:other_limits-challenges}.

In contrast, our analysis is based solely on the detection of high-energy neutrinos, without requiring an associated electromagnetic counterpart, thereby avoiding issues related to differences in the emission times of neutrinos and photons. Any stochastic spread in the emission times of neutrinos would tend to mask the possible LIV effects that we seek to constrain, weakening our results and rendering them more conservative. On the other hand, any correlation of neutrino energy with emission time could be confused with such an LIV effect, introducing a systematic uncertainty that cannot be quantified at the present stage.

Exploring LIV using neutrino data alone became feasible only following the IceCube observation of the 2014/2015 high-energy neutrino flare from TXS 0506+056, which consisted of multiple neutrinos with different energies.  Rather than comparing the arrival times of neutrinos and photons, we examine the distribution of arrival time of neutrinos of different energies in the flare.  This choice bypasses the complications associated to the modeling of multi-messenger emission from astrophysical sources. Nevertheless, our methods are not without challenges of their own, as we detail later.

In studying LIV in this way, we follow the suggestion for energetic photons made in \Refe~\cite{Amelino-Camelia:1997ieq}, but avoid subscribing to any one of the approaches mentioned in Sec.~\ref{sec:LVmotivate-overview}, thus lending general applicability to our methods.  Below, we show how violating Lorentz invariance would affect the time distribution of the neutrino flare.

%%%%%%%%%%%%%%%%%%%%%%%%%%%%%%%%%%%%%%%%%%%%%%%%%%%%%%%%%%%%%%%%%%%%%%%%%%%%
%%%%%%%%%%%%%%%%%%%%%%%%%%%%%%%%%%%%%%%%%%%%%%%%%%%%%%%%%%%%%%%%%%%%%%%%%%%%

\section{High-energy neutrino flares}
\label{sec:flares}

%%%%%%%%%%%%%%%%%%%%%%%%%%%%%%%%%%%%%%%%%%%%%%%%%%%%%%%%%%%%%%%%%%%%%%%%%%%%

\subsection{Sources of high-energy astrophysical neutrinos}
\label{sec:flares-sources}

In 2013, the IceCube neutrino telescope---the largest in operation---discovered TeV--PeV astrophysical neutrinos~\cite{IceCube:2013cdw, IceCube:2013low}.  Today, IceCube continues to observe them regularly~\cite{IceCube:2020wum, IceCube:2021uhz, IceCube:2024fxo}, and neutrino telescopes Baikal-GVD~\cite{Baikal-GVD:2022fmn, Baikal-GVD:2022fis} and KM3NeT~\cite{km3net_neutrino2024}, currently under construction, (and ANTARES~\cite{ANTARES:2024ihw}, now decommissioned) have reported hints of detection.
Still, the origin of the bulk of high-energy astrophysical neutrinos detected remains unknown~\cite{IceCube:2023oqe}.  They are likely produced predominantly by extragalactic sources---transient or steady-state---capable of accelerating particles up to energies of at least tens of PeV, and possibly much higher~\cite{AlvesBatista:2019tlv, AlvesBatista:2021eeu, Ackermann:2022rqc}.

So far, however, only a couple of likely candidate sources have been officially identified: transient emission from the blazar TXS 0506+056~\cite{IceCube:2018cha, IceCube:2018dnn}, about 1.75~Gpc away, and steady-state emission from the Seyfert galaxy NGC 1068~\cite{IceCube:2022der}, about 14~Mpc away.  Both have been associated to the detection of multiple neutrinos across a range of energies.  The significance of other candidate sources, typically identified based only on the detection of a single neutrino, is weaker; see Sec.~\ref{sec:other_limits} for an overview.  We expect that in the near future the number of sources will grow, owing to the ongoing and upcoming addition of more neutrino telescopes~\cite{Schumacher:2021hhm, Ackermann:2022rqc, Guepin:2022qpl}. (We do not consider high-energy neutrinos from the Galactic Plane~\cite{IceCube:2023ame, Bustamante:2023iyn} because they travel only a few kpc to us, making them less sensitive to the accumulation of LIV effects.)

Out of the above two likely candidate sources, only TXS 0506+056 has been observed to emit a flare of high-energy neutrinos, with a duration of a few months~\cite{IceCube:2018cha}; see \figu{event_weights}.   In contrast, the rate of detection of neutrinos from NGC 1068 was approximately uniform during its 10-year observation by IceCube~\cite{IceCube:2022der}, rendering it steady-state at least on that time scale.  Thus, we adopt TXS 0506+056 as the prototypical high-energy neutrino flare to illustrate our methods.

%%%%%%%%%%%%%%%%%%%%%%%%%%%%%%%%%%%%%%%%%%%%%%%%%%%%%%%%%%%%%%%%%%%%%%%%%%%%

\subsection{Challenges in detecting high-energy neutrinos}
\label{sec:flares-challenges}

There is significant uncertainty involved in inferring the energies of neutrinos detected from alleged astrophysical sources, like TXS 0506+056~\cite{IceCube:2013dkx, IceCube:2018cha}.  Because the size of the LIV-induced change in neutrino velocity (Secs.~\ref{sec:intro} and \ref{sec:LVmotivate}) depends on the neutrino energy, the uncertainty in neutrino energy is a limiting factor to the sensitivity of our analysis.  Below, we show how we account for this uncertainty  using realistic detection capabilities.

Astrophysical sources of high-energy neutrinos are primarily searched for by neutrino telescopes using muon tracks because they have pointing resolution of $1^\circ$ or better~\cite{Bradascio:2019eub, IceCube:2023oqe}.  These tracks are born in the charged-current deep inelastic scattering of high-energy astrophysical $\nu_\mu$ and $\bar{\nu}_\mu$ off nucleons, $N$, in the ice, \ie, $\nu_\mu + N \to \mu^- + X$ and its charge-conjugated process, where $X$ represents final-state hadrons.  The final-state hadrons receive a fraction $y$---the inelasticity---of the incoming neutrino energy, while the final-state muon receives the remaining fraction $(1-y)$.  The value of $y$ is random and unknown in each scattering; its distribution peaks at $y = 0$, but it has a wide spread.  At the TeV--PeV energies, the average inelasticity is about 0.25; see, \eg, \Refe~\cite{Weigel:2024gzh}.

The vast majority of the muons detected by IceCube are through-going, \ie, they are born in neutrino interactions that occur outside the instrumented detector volume~\cite{IceCube:2023oqe}.  As the final-state muon propagates through the ice it loses energy, leaving a track of Cherenkov light in its wake, only a segment of which crosses IceCube and deposits light in its photomultipliers.  From the direction of the track and the amount of light it deposits, IceCube uses simulations of $\nu_\mu$ interaction and muon propagation to infer the direction of the muon---which is closely aligned to that of its parent neutrino---and its \textit{energy proxy}, an estimate of the energy that the muon has as it enters the instrumented detector volume.  From this information, combined with the inelasticity distribution predicted by theory, analyses infer the most likely energy of the parent neutrino energy.

The above procedure, while valuable, has limitations.  Because the value of the inelasticity and the position where the through-going muon was born are unknown, there is significant uncertainty in the inferred neutrino energy.  In particular, the energy of the detected muon is only a lower limit on the energy of the parent neutrino.

%%%%%%%%%%%%%%%%%%%%%%%%%%%%%%%%%%%%%%%%%%%%%%%%%%%%%%%%%%%%%%%%%%%%%%%%%%%%

\subsection{Accounting for the neutrino energy uncertainty}
\label{sec:flares-uncertainty}

We adopt the IceCube detection of the 2014/2015 TXS 0506+056 neutrino flare as prototypical of the treatment of energy uncertainty, but the discussion applies generally to neutrino flares detected by any neutrino telescope via muon tracks.
We use the publicly released IceCube dataset~\cite{IceCube2018Dataset}, corresponding to the TXS 0506+056 analysis performed in~\cite{IceCube:2018cha}. This dataset provides the central values of the muon energy proxy, $\hat{E}_\mu$, for 320 events identified by IceCube as arriving from the declination of TXS 0506+056. These events were selected through a likelihood analysis that incorporated directional information during the IC86b observation period, which includes the time of the flare and is publicly available from~\cite{IceCube2018Dataset}.
\footnote{In our statistical analysis (Sec.~\ref{sec:liv_constraints_txs-mock_flares}), we later restrict the event sample to the subset occurring within the most likely flaring period.}

For each track, we find the distribution of the likely values of the parent neutrino energy, $E_\nu$.  We do this by accounting for two sources of energy uncertainty: the uncertainty in measuring the energy deposited by the muon track in the detector and the uncertainty in reconstructing the neutrino energy from the muon energy.  For the former, we assume that the muon energy is measured with a resolution of 10\% in $\log_{10}(\hat{E}_\mu/{\rm GeV})$, following IceCube prescriptions in \Refe~\cite{IceCube:2013dkx}.  For the latter, we use the relation between the energies of the muon and of its parent neutrino for each of the detected tracks, reported by IceCube in \Refe~\cite{IceCube:2018cha}.  The result is a probability density function, $\mathcal{P}_i(E_\nu)$ for the $i$-th detected track, that represents the most likely value of its parent neutrino energy and its spread.  Appendix~\ref{sec:pdf_neutrino_energy} contains the derivation of these functions and details on them.

%%%%%%%%%%%%%%%%%%%%%%%%%%%%%%%%%%%%%%%%%%%%%%%%%%%%%%%%%%%%%%%%%%%%%%%%%%%%
\begin{figure}[t!]
 \centering
 \includegraphics[width=1.0\columnwidth]{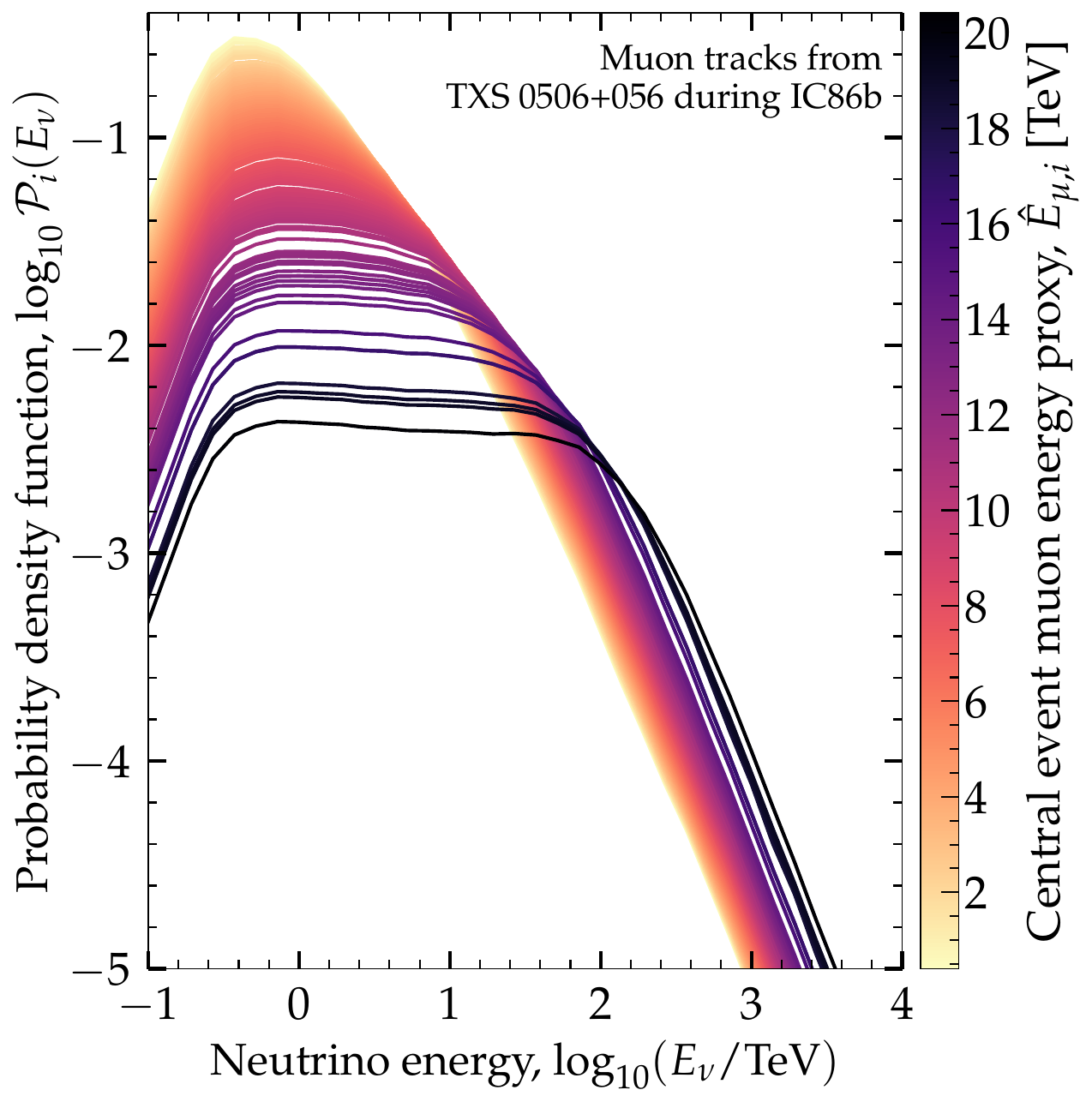}
 \caption{\label{fig:prob_energy_nu}\textbf{\textit{Probability density function of the neutrino energy, $E_\nu$.}} Each curve represents a different muon track detected by IceCube during the IC86b period that arrived from the direction of TXS 0506+056. The events are characterized by their central reconstructed muon energy proxy, $\hat{E}_{\mu,i}$. For each event, the PDF is constructed by incorporating the relationship between the true neutrino energy and the muon energy proxy, as well as the correspondence between this quantity and the observed muon energy proxy. See Sec.~\ref{sec:flares-uncertainty} and Appendix~\ref{sec:pdf_neutrino_energy} for further details.}
\end{figure}
%%%%%%%%%%%%%%%%%%%%%%%%%%%%%%%%%%%%%%%%%%%%%%%%%%%%%%%%%%%%%%%%%%%%%%%%%%%%

Figure~\ref{fig:prob_energy_nu} shows these probability density functions for all the IceCube events detected during the 2014/2015 TXS 0506+056 flare. For tracks with a low muon energy proxy, of roughly 0.1--1~TeV, the most likely value of the parent neutrino energy is similar.  For tracks with a higher muon energy proxy, of roughly 1--100~TeV, the distribution of parent neutrino energies is flatter and reaches higher energies, reflecting the fact that, for higher-energy neutrinos, there are more combinations of the inelasticity and the position of the $\nu_\mu N$ interaction that can yield a high-energy track.

Later (Sec.~\ref{sec:liv_constraints_txs-mock_flares}), we use these functions to
sample randomly neutrino energies and generate mock neutrino flares, which we then use to probe LIV.

%%%%%%%%%%%%%%%%%%%%%%%%%%%%%%%%%%%%%%%%%%%%%%%%%%%%%%%%%%%%%%%%%%%%%%%%%%%%
%%%%%%%%%%%%%%%%%%%%%%%%%%%%%%%%%%%%%%%%%%%%%%%%%%%%%%%%%%%%%%%%%%%%%%%%%%%%

\section{Tests of Lorentz invariance in high-energy neutrino flares}
\label{sec:methods}

%%%%%%%%%%%%%%%%%%%%%%%%%%%%%%%%%%%%%%%%%%%%%%%%%%%%%%%%%%%%%%%%%%%%%%%%%%%%

\subsection{Time delays in neutrino propagation}
\label{sec:methods-delays}

As was done for photons in \Refe~\cite{Ellis:2018lca},
we consider LIV-induced modifications of the neutrino velocity of the form
\begin{equation}
 \label{equ:vg1}
 v(E_\nu)
 =
 \left[1 - \frac{n+1}{2}\left(\frac{E_\nu}{M_n}\right)^n\right] %\ \;,
 \equiv
 1 - \Delta v(E_\nu) \;,
\end{equation}
where $M_n$ is the LIV energy scale, $E_\nu$ is the neutrino energy, and $\Delta v$ is the deviation in the neutrino velocity relative to the speed of light. Here and below we use natural units where the speed of light $c = 1$.

Since we are interested in neutrinos from sources located at substantial
redshifts, we must account for the change to the neutrino energy
due to the cosmological expansion.  An astrophysical neutrino detected on Earth, at $z = 0$, with
energy $E_\nu$, had an energy $E_\nu(1+z)$ at redshift $z$.
We adopt the standard $\Lambda$CDM cosmological model,
in which the relation between time, $t$, and redshift, $z$, is
$dt=-dz [H_0(1+z)h(z)]^{-1}$, where
$h(z) = \sqrt{\Omega_{\Lambda} + \Omega_m(1+z)^3}$, $H_0 = 70$~km~s$^{-1}$~Mpc$^{-1}$
is the Hubble constant, and $\Omega_\Lambda = 0.69$ and $\Omega_m = 0.31$ are, respectively
the dimensionless energy density fractions of the vacuum and matter~\cite{Planck:2018vyg}.\footnote{The uncertainties on these cosmological parameters can be neglected for our purposes.}

Under LIV, the proper travel distance of a neutrino differs from its travel distance at light-speed.  LIV induces a differential change in the travel distance
of size $d(\Delta L) = \Delta v~dt$, where $\Delta v$ is from \equ{vg1}.
Upon reaching Earth, a neutrino emitted from a source at redshift $z_{\rm src}$ has accumulated
a total path difference, $\Delta L_n$.  Because we deal with relativistic neutrinos, the delay
in the arrival time of the neutrino at Earth is $\Delta t_n \approx \Delta L_n$, \ie,
\begin{equation}
 \label{equ:delta_time}
 \Delta t_n
 =
 H_0^{-1}
 \int_{0}^{z_{\rm src}}
 \frac{\left. \Delta v\right\vert_{E_\nu(1+z)}}{h(z)} dz
 \equiv
 \tau_n(z_{\rm src}) E_\nu^n
 \;,
\end{equation}
where the integrand is evaluated at $E_\nu(1+z)$ and
\begin{equation}
 \label{equ:tauK1}
 \tau_n(z_{\rm src}) = a_n K_n(z_{\rm src}) \;,
\end{equation}
with $a_n\equiv\frac{n+1}{2}\frac{H_0^{-1}}{M_n^n}$ and
$K_n(z_{\rm src})=\int_0^{z_{\rm src}}\frac{(1+z)^n}{h(z)} dz$.  As a result, the difference in arrival times between two neutrinos detected with energies $E_{\nu, 2} > E_{\nu, 1}$ is
$\tau_n (E_{\nu, 2}^n - E_{\nu, 1}^n)$.

Since the production mechanism of high-energy astrophysical neutrinos is presently unknown~\cite{AlvesBatista:2019tlv, AlvesBatista:2021eeu, Ackermann:2022rqc},
we allow for the possibility that, in a neutrino flare, neutrinos of different energies are emitted
at different times.  To do this, we consider a potential energy-dependent
time-lag at the source, $b_{\rm s}$.  Thus, under LIV, the arrival time of a neutrino upon reaching Earth is
\begin{equation}
 t_{\rm obs}(E_\nu)
 =
 b_{\rm s}(E_\nu)(1+z_{\rm src})
 +
 \tau_{n}(z_{\rm src})E_\nu^n \;.
 \label{equ:lag1}
\end{equation}

The standard expectation, \ie, in the absence of LIV, is that the time distribution of neutrinos in a flare reaching the Earth has the same shape as when it was emitted by the source; see \figu{overview}. (The number of neutrinos in the flare is, of course, lower at Earth than near the source.)  Thus, we need to determine how much the neutrino time distribution at Earth under LIV differs from the standard expectation.  In an ideal, but unrealistic scenario where we knew what was the time distribution at emission time, this comparison would be straightforward.

In reality, however, we do not know the emitted neutrino time distribution, since it depends on the largely uncertain physical conditions inside the candidate astrophysical sources.  On top of that, any one observed flare is subject to the internal, unknown stochasticity of the mechanism that produced it.  (Added to that, there is the experimental uncertainty on the neutrino energy; see Secs.~\ref{sec:flares-challenges} and \ref{sec:flares-uncertainty}.)  This presents a major challenge in identifying the presence of LIV in a neutrino flare.

We surmount this challenge below (Sec.~\ref{sec:methods-measures}) by introducing estimators of the irregularity, kurtosis, and skewness of the neutrino time distribution in a flare, borrowed from \Refe~\cite{Ellis:2018lca}, that can signal the presence of generic LIV-induced features in the detected time distribution, regardless of what it was at emission.  Before, however, we introduce \textit{event weights}, which we use in computing the estimators.

%%%%%%%%%%%%%%%%%%%%%%%%%%%%%%%%%%%%%%%%%%%%%%%%%%%%%%%%%%%%%%%%%%%%%%%%%%%%

\subsection{Energy and directional event weights in a flare}
\label{sec:methods-weights}

In the sample of events---muon tracks---detected by IceCube during the 2014/2015 TXS 0506+056 neutrino flare~\cite{IceCube:2018cha, IceCube2018Dataset}, the $i$-th detected event has arrival time $t_{{\rm obs},i}$, direction given by declination $\theta_{z,i}$ and right ascension $\phi_{z,i}$, with angular uncertainty $\sigma_{\Omega_{z,i}}$, and central value of the muon energy proxy $\hat{E}_{\mu ,i}$.  In developing our methods, we assume that this information would also be available for any future observed neutrino flare.

From this information, we compute two weights for each event: an energy weight, $w_{{E_\nu},i}$, and a directional weight, $w_{{\Omega_z},i}$. They serve two purposes: first, they allow us to prioritize events originating from directions closer to the candidate neutrino source by assigning them greater weights in the computation of the estimators, and second, to emphasize events with higher energies, as higher-energy neutrinos are more likely to be of astrophysical rather than atmospheric origin.

%%%%%%%%%%%%%%%%%%%%%%%%%%%%%%%%%%%%%%%%%%%%%%%%%%%%%%%%%%%%%%%%%%%%%%%%%%%%
\begin{figure}[t!]
 \centering
 \includegraphics[width=0.95\columnwidth, trim={0 0 2cm 0, clip}]{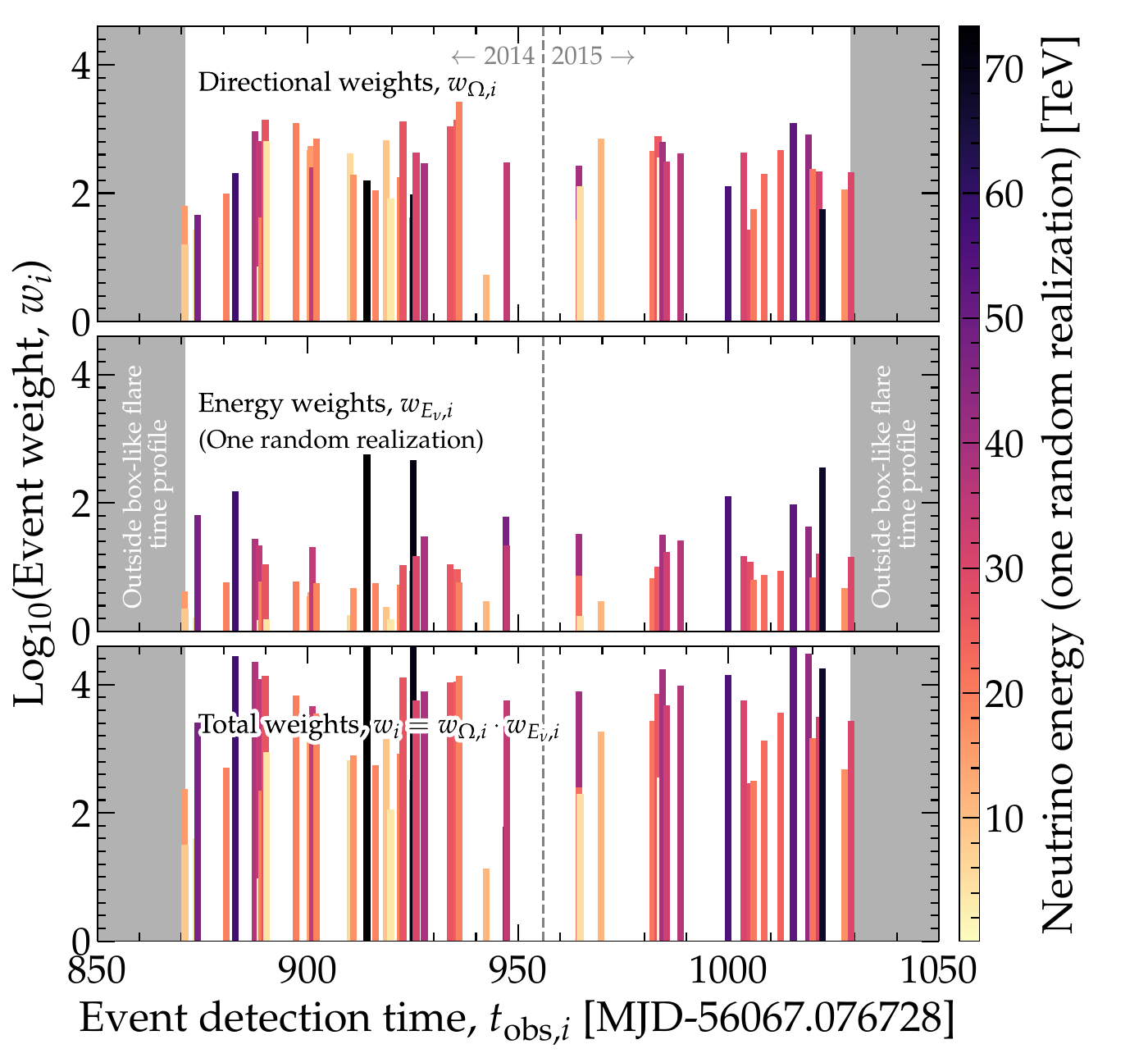}
 \caption{\label{fig:event_weights}
 \textbf{\textit{Analysis weights associated to neutrinos detected from a mock flare.}}  The neutrino energies in this illustrative mock flare were randomly drawn from their probability distributions (Sec.~\ref{sec:liv_constraints_txs-mock_flares}).  Our statistical analysis is based on generating many such mock flares. \textit{Top:} Directional weights, \equ{weight_direction}; events from directions closer to TXS 0506+056 and with better angular resolution have higher weights.  \textit{Middle:}
 Energy weights, \equ{weight_energy}, shown here for one random mock neutrino flare.  Neutrinos with higher energy have higher weights.  \textit{Bottom:}  Total weights, obtained by multiplying the directional and energy weights.  Neutrinos with higher weights are more likely to come from the TXS 0506+056 flare.  See Sec.~\ref{sec:liv_constraints_txs-mock_flares} for details.}
\end{figure}
%%%%%%%%%%%%%%%%%%%%%%%%%%%%%%%%%%%%%%%%%%%%%%%%%%%%%%%%%%%%%%%%%%%%%%%%%%%%

Figure~\ref{fig:event_weights} shows, for illustration, the weights for one random mock flare, modeled after the 2014/2015 TXS 0506+056 flare, containing $N_\nu$ events.  We introduce random mock flares as part of our statistical analysis later, in Sec.~\ref{sec:liv_constraints_txs}.  For each mock detected event, we randomly sample the energy of its parent neutrino, $E_{\nu, i}$, from its probability density function, $\mathcal{P}_i$ (Sec.~\ref{sec:flares-uncertainty}).

\textbf{\textit{Directional weight.---}}The directional weight of an event measures the angular distance between its arrival direction and the position of TXS 0506+056.  The angular distance is computed using the Haversine formula, \ie, $\Delta\Omega_i = {\rm hav}(\theta_{z,i}-\theta_{z,{\rm src}}) + \cos(\theta_{z,i}) \cos(\theta_{z,{\rm src}}) {\rm hav}(\phi_{z,i}-\phi_{z,{\rm src}})$, where ${\rm hav}(\theta) \equiv \sin^2(\theta/2)$.  The directional weights are then computed as
\begin{equation}
 \label{equ:weight_direction}
 w_{\Omega,i}
 =
 \frac{\Delta\Omega_i}{{\rm min} \left\{ \Delta\Omega_j \right\}_{j=1}^{N_\nu}} \;,
\end{equation}
where the denominator denotes the minimum separation from the source position among the events.

\smallskip

\textbf{\textit{Energy weight.---}}The energy weight of an event is computed in analogy to what was done for gamma rays in \Refe~\cite{Ellis:2018lca}, \ie,
\begin{equation}
 \label{equ:weight_energy}
 w_{E_\nu,i}
 =
 \frac{E_{\nu,i}}{{\rm min} \left\{ E_{\nu,j} \right\}_{j=1}^{N_\nu}} \;,
\end{equation}
where the denominator denotes the minimum neutrino energy among the events.

\smallskip

\textbf{\textit{Total weight.---}}The total weight of the event is the product of the energy and directional weights, \ie,
\begin{equation}
 \label{equ:weight_total}
 w_i = w_{E_\nu,i} \cdot w_{\Omega,i} \;.
\end{equation}
The absolute values of the weights are irrelevant; what matters are their relative values, both among individual events and across different mock flares.

\smallskip

Our calculation of weights is akin to that used by the IceCube Collaboration to discover the neutrino flare~\cite{Braun:2008bg, IceCube:2015usw, IceCube:2018cha} as a means to quantify the spatial and temporal clustering of events using an unbinned maximum likelihood function.  The significance of the association of the flare with an astrophysical source is driven by the events with large weights.

%%%%%%%%%%%%%%%%%%%%%%%%%%%%%%%%%%%%%%%%%%%%%%%%%%%%%%%%%%%%%%%%%%%%%%%%%%%%

\subsection{Background atmospheric neutrinos and muons}
\label{sec:methods-background}

When searching for astrophysical sources of high-energy neutrinos~\cite{IceCube:2023oqe, IceCube:2023myz}, there is an irreducible contamination from the high background of atmospheric neutrinos and muons created in the interactions of cosmic rays in the atmosphere of the Earth.  Short-lived candidate sources---like GRBs, whose prompt emission lasts only seconds---are largely impervious to this background, but longer-lived sources---like a flaring blazar, which may be active for months---have a sizable estimated atmospheric contamination.  For instance, it is estimated~\cite{IceCube:2018cha} that only $13 \pm 5$ events detected during the 2012--2015 period of observation of TXS 0506+056 were due to neutrinos, and the rest to atmospheric muons.

The simulation of the propagation of high-energy muons through matter and the detection of their Cherenkov radiation by the photomultipliers of in-ice and in-water neutrino telescopes is a complex, computationally taxing task that requires in-depth knowledge of the detector; see \Refes~\cite{Chirkin:2004hz, Koehne:2013gpa} for how IceCube and KM3NeT do it.  Due to this complexity---best left for an internal analysis by experimental collaborations---we do not explicitly model the presence of the atmospheric neutrino and muon backgrounds in our search for LIV.

Instead, we mitigate the impact of atmospheric backgrounds in our analysis by applying the directional and energy weights introduced earlier. Events with higher total weights, \equ{weight_total}, are more likely to be astrophysical rather than atmospheric in origin. The LIV estimators we introduce below are designed to prioritize events with higher weights, thereby enhancing the influence of astrophysical neutrinos in the flare and reducing the impact of atmospheric neutrinos.

%%%%%%%%%%%%%%%%%%%%%%%%%%%%%%%%%%%%%%%%%%%%%%%%%%%%%%%%%%%%%%%%%%%%%%%%%%%%

\subsection{Estimators of LIV in a flare time distribution}
\label{sec:methods-measures}

From \equ{lag1}, we anticipate that LIV would deform the time distribution of neutrinos in a flare~\cite{Ellis:2018lca}.
To probe the presence of LIV in the pattern of neutrino arrival times,
we rely on times measured in the detector reference frame, given by
\begin{equation}
 b_{{\rm d},i}(\tau_n) = t_{{\rm obs},i} - \tau_n (z_{\rm src}) E_{\nu, i}^n \; ,
 \label{equ:rec1}
\end{equation}
where the compensation parameter, $\tau_n$, is taken from \equ{tauK1}.
Thus, the original pattern of intrinsic emission times, $b_{{\rm s}, i}$ in \equ{lag1} (there expressed in the source frame),
can, in principle, be recovered by subtracting the appropriate values of $\tau_n$
from the observed arrival times, so as to compensate the impact of the
LIV-induced dispersion~\cite{Ellis:2018lca}.

\begin{figure}[t!]
 \centering
 \includegraphics[width=1.0\columnwidth]{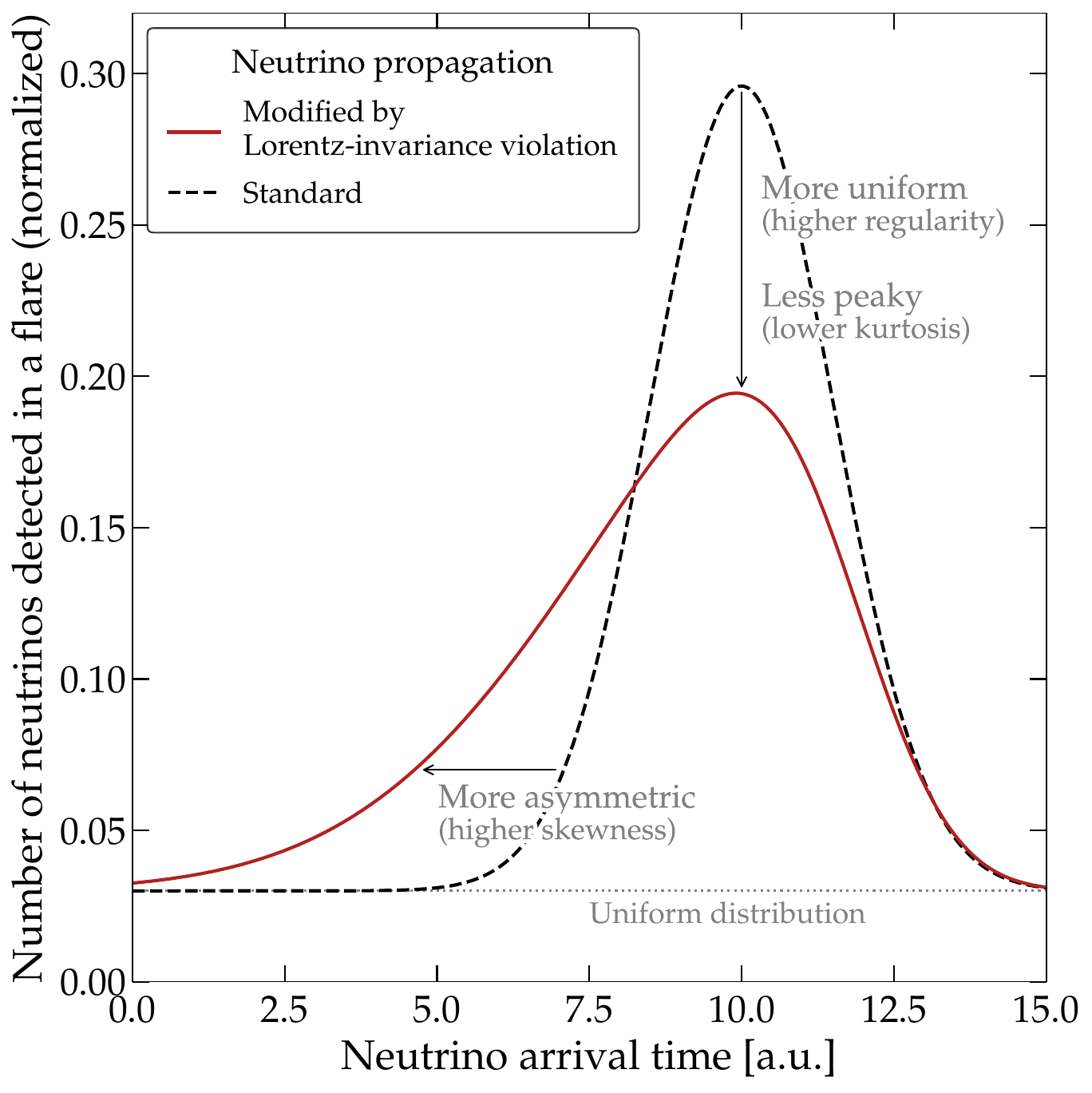}
 \caption{\label{fig:time_profile}
 \textbf{\textit{Sketch of the effects of Lorentz-invariance violation on the arrival times of neutrinos in a flare.}} In this figure we assume---for illustration only---that, in the absence of LIV during neutrino propagation, the arrival times of high-energy neutrinos at Earth would be distributed normally, preserving the time distribution emitted by their astrophysical source.  Yet, in our calculations, we do not assume any shape for the emission time profile.  Under LIV, the distribution of arrival times may deviate from the distribution at emission, and become more uniform, less peaky, and more asymmetric.  See Sec.~\ref{sec:methods-measures} for details.}
\end{figure}

Figure~\ref{fig:time_profile} sketches the effects of LIV on the time distribution of the neutrinos in a flare: regardless of the original shape of the distribution, the LIV-induced dispersion makes it more uniform, less peaky, and more asymmetric. Following the approach introduced in \Refe~\cite{Ellis:2018lca}, we consider these
three distinct deformations and introduce below non-parametric statistical estimators to quantify them. Their purpose is to find, in three complementary manners, the value of $\tau_n$ that rectifies the above LIV-induced deformations of the time distribution---\ie, that undoes the LIV effects.  From this, we estimate the size of LIV itself; specifically, of its energy scale, $M_n$ in \equ{vg1}.

%%%%%%%%%%%%%%%%%%%%%%%%%%%%%%%%%%%%%%%%%%%%%%%%%%%%%%%%%%%%%%%%%%%%%%%%%%%%
%%%%%%%%%%%%%%%%%%%%%%%%%%%%%%%%%%%%%%%%%%%%%%%%%%%%%%%%%%%%%%%%%%%%%%%%%%%%

\smallskip

\textbf{\textit{Irregularity.---}}In the presence of LIV,
the time distribution of a neutrino flare undergoes dispersion,
leading to the temporal spread of neutrinos with different energies as they propagate.
Eventually, this leads to the blending of any initially distinct distribution with a uniform background, regardless of the shape of the initial flare.

Given a mock flare (Sec.~\ref{sec:liv_constraints_txs-mock_flares}) and a trial value of $\tau_n$, we compute the intrinsic
times, $b_{{\rm d}, i}$ in \equ{rec1}, where $i = 1, \dots, N_\nu$, and $N_\nu$ is the number of detected neutrinos.  With them, we build the cumulative time distribution, $\Xi(\tau_n, b_{\rm d})$, as the sum of the weights of neutrinos detected starting from the earliest value of $b_{{\rm d}, i}$ in the flare---\ie, the first neutrino detected---up to the time $b_{\rm d}$, and divided by the sum of all weights in the flare.  We compare this with the cumulative distribution $U(\tau_n, b_{\rm d})$ built assuming that the values of $b_{\rm d}$ are distributed uniformly through the duration of the flare, \ie, between the earliest and latest values of $b_{\rm d}$.

If neutrinos in the flare
were affected by LIV en route to Earth, one specific value of
$\tau_n$ should best compensate the arrival times in such a way as to reconstruct the original, irregular
time distribution of the flare.  The degree of irregularity is estimated using the Kolmogorov-Smirnov (KS) statistic~\cite{Ellis:2018lca}, \ie,
\begin{equation}
 \label{equ:estimator_ks}
 \mathcal{D}(\tau_n)
 =
 \max_{b_{{\rm d}, 1} \leq b_{\rm d} \leq b_{{\rm d}, N_\nu}}
 \left\vert
 \Xi(\tau_n, b_{\rm d})
 -
 U(\tau_n, b_{\rm d})
 \right \vert \;.
\end{equation}
For a given mock flare, our measure of LIV is $\tau_n^{\rm KS}$, the value of the compensation parameter that maximizes $D$.

\smallskip

\textbf{\textit{Kurtosis.---}}Regardless of the initial shape of the
time distribution of the neutrino flare, LIV en route to Earth may alter its kurtosis in a distinct, energy-dependent manner. Specifically, it makes a flare evolve towards a flattened, or platykurtic (negative kurtosis) distribution.  Consequently, we adopt a compensation procedure that restores the distribution
to its maximally peaky shape.

For a given trial value of $\tau_n$, we compute the excess kurtosis---compared to a normal distribution---of the time distribution of the neutrino flare as
\begin{equation}
 \label{equ:estimator_kurt}
 \mathcal{K}(\tau_n)
 =
 N_W
 \frac{
 \sum_{i=1}^{N_\nu}
 \left\{
 \left[
 b_{{\rm d},i}(\tau_n) - \bar{b}_{\rm d}(\tau_n)
 \right]
 w_i
 \right\}^4
 }
 {
 \left(
 \sum_{i=1}^{N_\nu}
 \left\{
 \left[
 b_{{\rm d},i}(\tau_n) - \bar{b}_{\rm d}(\tau_n)
 \right]
 w_i
 \right\}^2
 \right)^2
 }
 -
 3 \;,
\end{equation}
where $w_i$ is the total event weight from \equ{weight_total}, $\bar{b}_{\rm d}(\tau_n) \equiv N_\nu^{-1} \sum_{i=1}^{N_\nu} b_{{\rm d},i}(\tau_n)$ is the average value of the times and $N_W$ is a normalization constant whose value is unimportant (we set $N_W = 1$), since we only care about differences in kurtosis between different random mock flares (Sec.~\ref{sec:liv_constraints_txs}).
For a given mock flare, our measure of LIV is $\tau_n^{\rm kurt}$, the value of the compensation parameter that maximizes $\mathcal{K}$.

\smallskip

%%%%%%%%%%%%%%%%%%%%%%%%%%%%%%%%%%%%%%%%%%%%%%%%%%%%%%%%%%%%%%%%%%%%%%%%%%%%
\begin{figure}[t!]
 \centering
 \includegraphics[width=0.95\columnwidth]{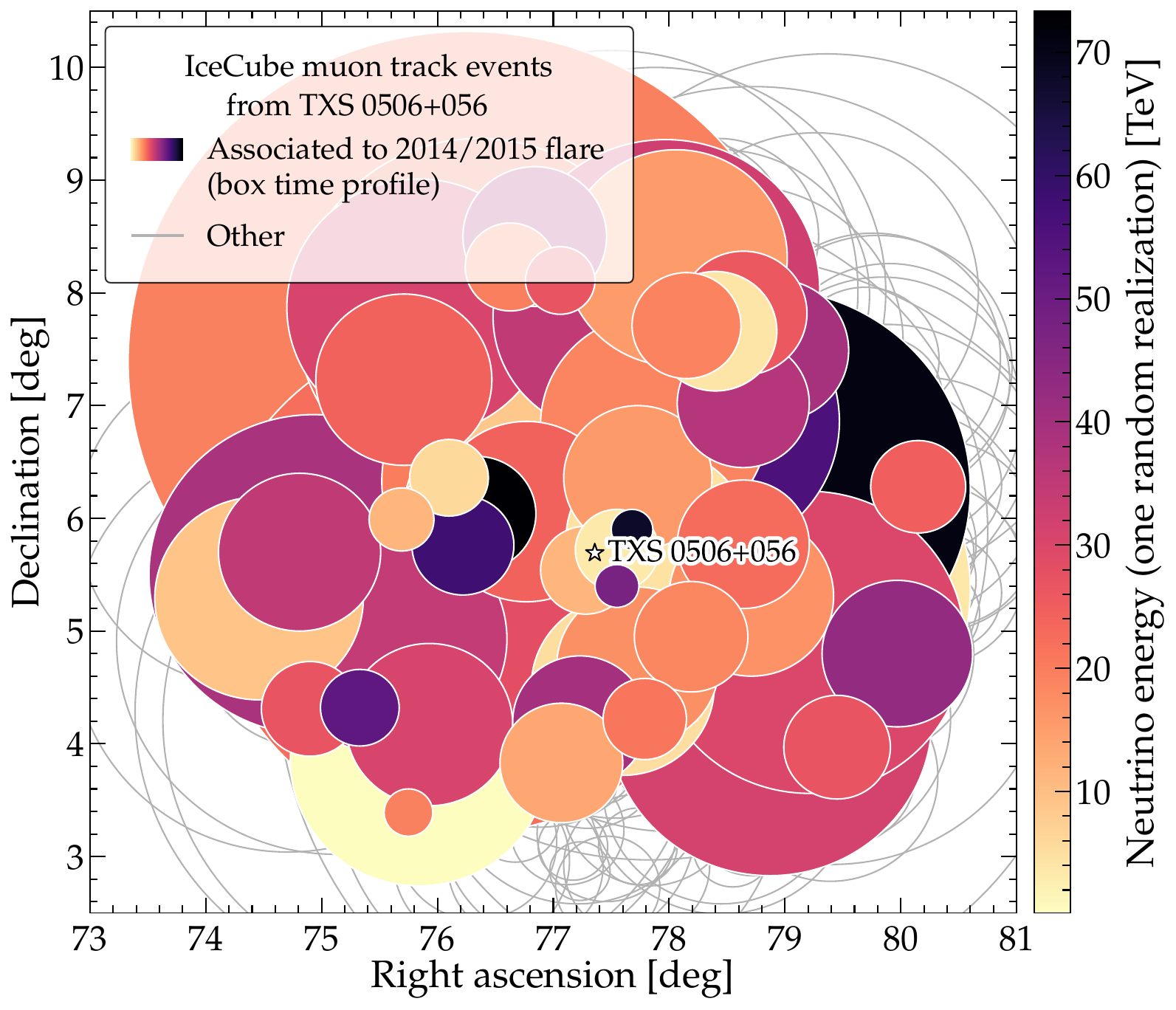}
 \caption{\label{fig:event_positions}
 \textbf{\textit{Directions of the IceCube events detected during the 2014/2015 TXS 0506+056 flare.}} All events are  muon tracks.  For each event, its 68\%~C.L.~angular uncertainty is reflected by the size of its corresponding disc. In our analysis, we use only the events most closely associated to the flare, \ie, those resulting from the box time profile from the IceCube analysis in \Refe~\cite{IceCube:2018cha}.  See Sec.~\ref{sec:liv_constraints_txs-mock_flares} for details.}
\end{figure}
%%%%%%%%%%%%%%%%%%%%%%%%%%%%%%%%%%%%%%%%%%%%%%%%%%%%%%%%%%%%%%%%%%%%%%%%%%%%

\textbf{\textit{Skewness.---}}Also regardless of the initial shape of the time distribution of the neutrino flare, LIV en route to Earth may increase its asymmetry, which can be measured via its skewness.
LIV-induced dispersion tends to make the skewness of the time distribution negative, compared to a normal distribution~\cite{Ellis:2018lca}. Conversely,
our compensation procedure aims to undo the effects of LIV by shifting the skewness towards positive values.

For a given trial value of $\tau_n$, we compute the skewness of the time distribution of the neutrino flare as
\begin{equation}
 \label{equ:estimator_skew}
 \mathcal{S}(\tau_n)
 =
 \sqrt{N_W}
 \frac{
 \sum_{i=1}^{N_\nu}
 \left\{
 \left[
 b_{{\rm d},i}(\tau_n) - \bar{b}_{\rm d}(\tau_n)
 \right]
 w_i
 \right\}^3
 }
 {
 \left(
 \sum_{i=1}^{N_\nu}
 \left\{
 \left[
 b_{{\rm d},i}(\tau_n) - \bar{b}_{\rm d}(\tau_n)
 \right]
 w_i
 \right\}^2
 \right)^{3/2}
 } \;,
\end{equation}
where $\bar{b}_{\rm d}$ is defined as before and, again, we set $N_W = 1$.  For a given mock flare, our measure of LIV is $\tau_n^{\rm skew}$, the value of the compensation parameter that maximizes $\mathcal{S}$.

\medskip

When estimating the possible size of LIV effects, we do so using separately the three estimators above. We report their results in Sec.~\ref{sec:results}.

%%%%%%%%%%%%%%%%%%%%%%%%%%%%%%%%%%%%%%%%%%%%%%%%%%%%%%%%%%%%%%%%%%%%%%%%%%%%
%%%%%%%%%%%%%%%%%%%%%%%%%%%%%%%%%%%%%%%%%%%%%%%%%%%%%%%%%%%%%%%%%%%%%%%%%%%%

\section{LIV constraints from the TXS 0506+056 neutrino flare}
\label{sec:liv_constraints_txs}

%%%%%%%%%%%%%%%%%%%%%%%%%%%%%%%%%%%%%%%%%%%%%%%%%%%%%%%%%%%%%%%%%%%%%%%%%%%%
\begin{figure*}[t!]
 \centering
 \includegraphics[width=1.0\columnwidth]{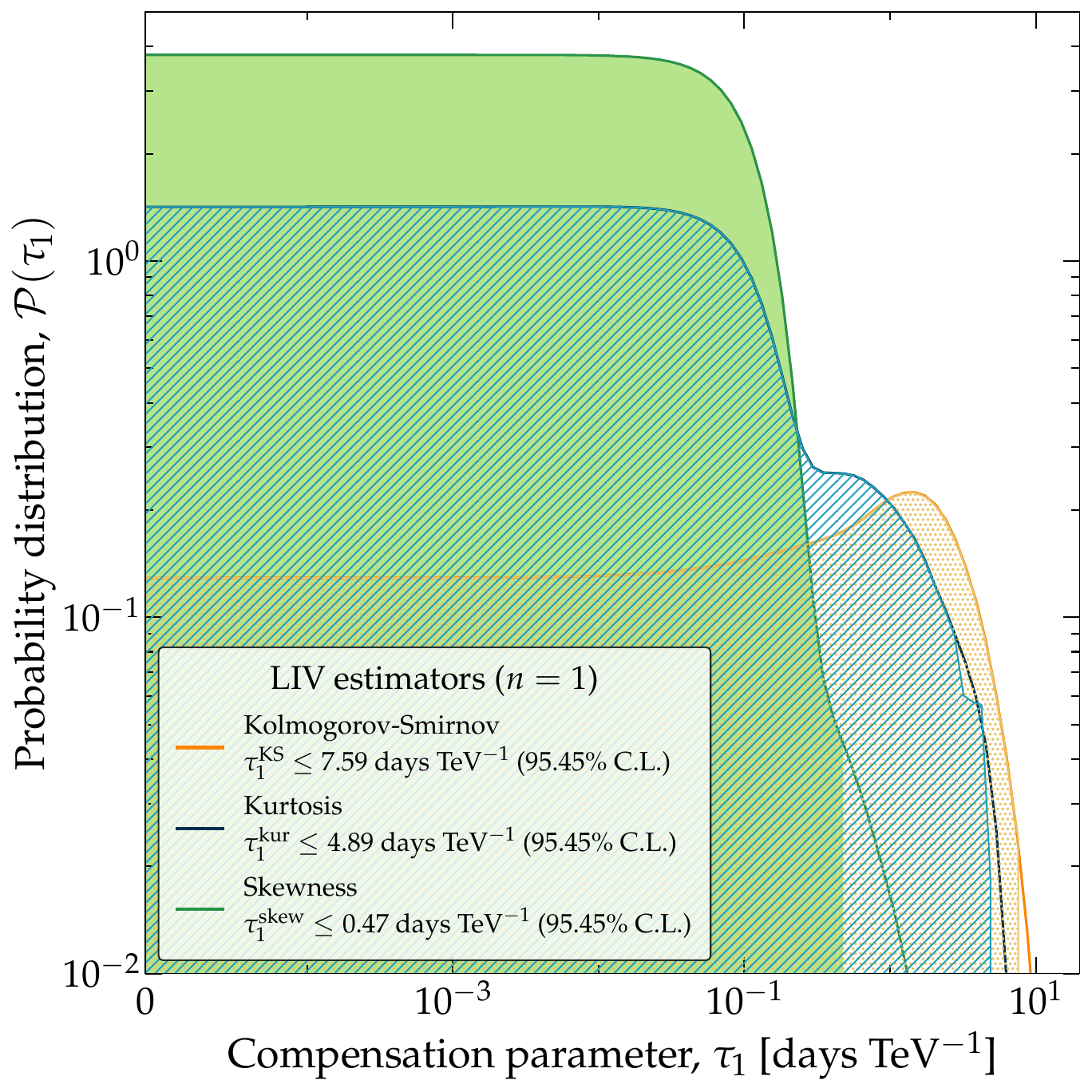}
 \includegraphics[width=1.0\columnwidth]{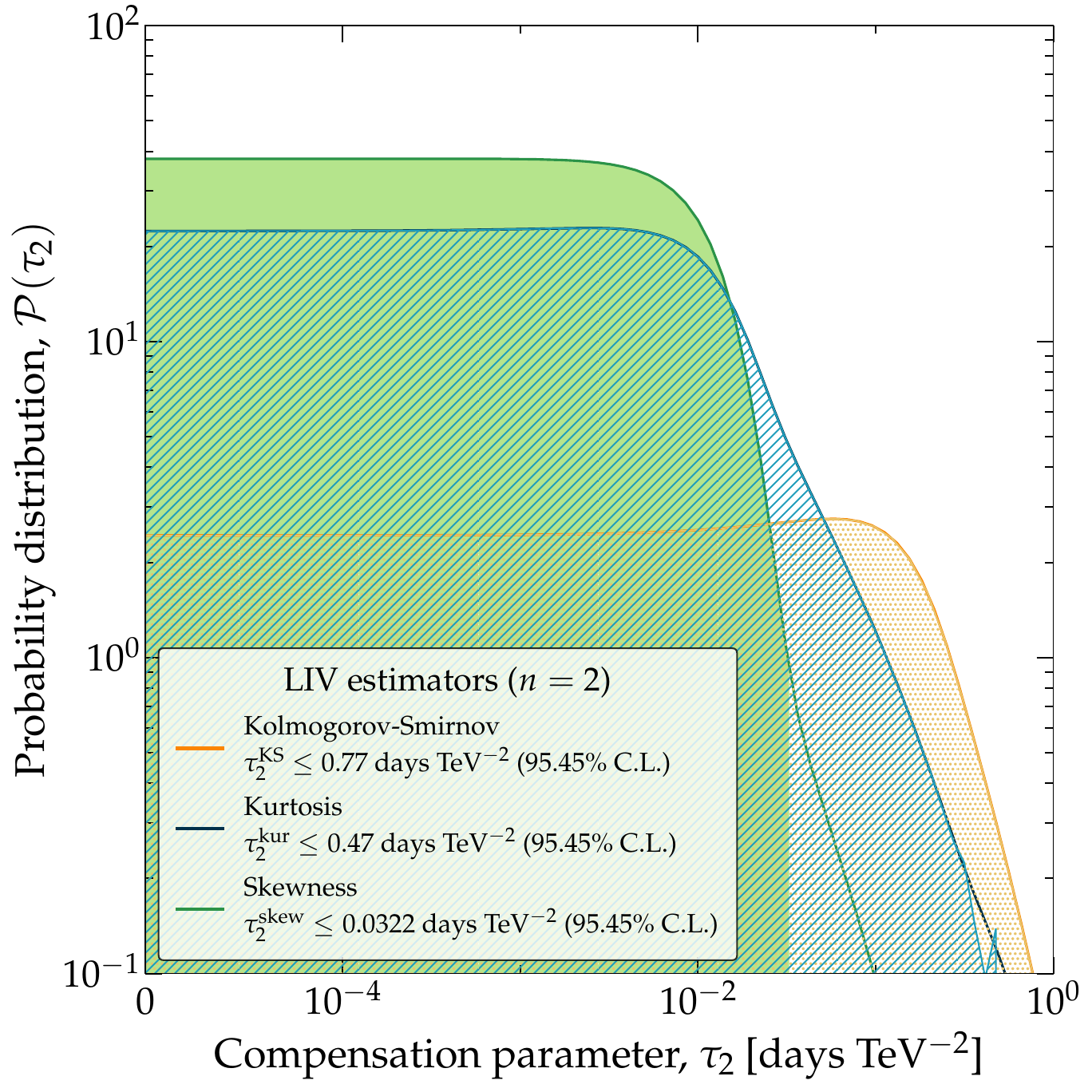}
 \caption{\label{fig:estimator_distributions}
 \textbf{\textit{Probability distributions of the Lorentz-invariance-violation estimators in neutrino propagation.}}  The distributions are inferred from examining the shape of the time distribution of the high-energy neutrinos detected from the 2014/2015 flare of the blazar TXS 0506+056, detected by IceCube.  We use three estimators to measure deviations from the standard expectation for the time distribution: Kolmogorov-Smirnov---a measure of its regularity---kurtosis, and skewness.  The strength of the LIV effects grows with the neutrino energy, \ie, $\propto E^{n}$.
 \textit{Left:} Assuming that the effects are linear in neutrino energy, \ie, $\propto E^{n}$, with $n = 1$.  \textit{Right:} Assuming that they are quadratic in neutrino energy, \ie, $n = 2$.  All the limits shown are at 95\%~C.L.  The distributions are shown after applying a Gaussian kernel density estimator to smooth them.  See Secs.~\ref{sec:methods} and \ref{sec:liv_constraints_txs} for details, and Eqs.~(\ref{equ:limit_m1_ks})--(\ref{equ:limit_m2_skew}) for the resulting limits on the energy scale of LIV.
 The distributions are one-sided, \ie, they assume that $\tau_n \geq 0$, implying that LIV slows down neutrinos.
 \textbf{\textit{These limits rely only on the detection of a flare of high-energy neutrinos from an astrophysical source, and do not require the coincident detection of an electromagnetic flare from the same source.}}}
\end{figure*}
%%%%%%%%%%%%%%%%%%%%%%%%%%%%%%%%%%%%%%%%%%%%%%%%%%%%%%%%%%%%%%%%%%%%%%%%%%%%

To address whether there is evidence of LIV in the time distribution of the neutrinos from a detected flare, we build---via mock flares based on it, as we explain below---the distribution of allowed values of the LIV measures (Sec.~\ref{sec:methods-measures}), $\tau_n^{\rm KS}$, $\tau_n^{\rm kurt}$, and $\tau_n^{\rm skew}$.

%%%%%%%%%%%%%%%%%%%%%%%%%%%%%%%%%%%%%%%%%%%%%%%%%%%%%%%%%%%%%%%%%%%%%%%%%%%%

\subsection{Mock neutrino flares}
\label{sec:liv_constraints_txs-mock_flares}

We base our mock neutrino flares on a specific detected flare.  Like before, we use as prototype for the latter the 2014/2015 TXS 0506+056 flare detected by IceCube, but our methods can be applied to similar future flares.

Figure~\ref{fig:event_positions} shows the arrival directions of the IceCube events that we use in our analysis, compared to the position of TXS 0506+056.  We use only the events identified in the IceCube analysis of~\cite{IceCube:2018cha} as originating from the general direction of TXS 0506+056 during the 158-day period spanning modified Julian dates 56937.81 to 57096.21. This period contains $N_\nu = 58$ events, corresponding to the time window defined by the simple box-shaped time profile used in~\cite{IceCube:2018cha} to characterize the neutrino flare.\footnote{A Gaussian time profile~\cite{IceCube:2018cha} would yield a similar selection of events.} We refer to this set as a ``flare-containing sample", i.e., a subset of the data likely to include the flare reported in~\cite{IceCube:2018cha} but not necessarily limited to it. Using this flare-containing sample, rather than the full dataset analyzed in~\cite{IceCube:2018cha}, which covers several years of observation, helps reduce contamination from unrelated background events. From this restricted sample, we construct mock neutrino flares by generating random samples, as described below. Since our analysis employs non-parametric estimators (as distinct from fit parameters), we do not perform a time-profile flare-defining analysis as done in~\cite{IceCube:2018cha}. Instead, each mock flare is associated with the full multiplicity (58) of the selected subset.

In all mock flares, the arrival time, $t_{{\rm obs},i}$, direction, $\theta_{z,i}$ and $\phi_{z,i}$, and angular uncertainty, $\sigma_{\theta_{z, i}}$, of the $i$-th mock event are the same as that of $i$-th event in the detected flare.  For the arrival times, this is justified because arrival times are measured with nanosecond precision by IceCube, which leaves essentially no wiggle room.  For the arrival directions, while we fix the best-fit direction of the event, we account for its directional uncertainty via the directional weights, \equ{weight_direction}.  What varies in each random mock flare is the neutrino energy, $E_{\nu, i}$, associated to each event: for the $i$-th event its value is randomly sampled from the probability density function $\mathcal{P}_i(E_\nu)$ (Sec.~\ref{sec:flares-challenges}).

Thus, the $i$-th event in a random mock flare is characterized by $e_i \equiv \left( t_{{\rm obs},i}, \theta_{z,i}, \phi_{z,i}, \sigma_{\Omega_{z,i}}, E_{\nu, i} \right)$, and the flare is made up of the collection of mock events $\left\{ e_1, \ldots e_{N_\nu} \right\}$.  For each event we compute the directional weight, $w_{{\Omega_z},i}$ [\equ{weight_direction}], energy weight, $w_{{E_\nu},i}$ [\equ{weight_energy}], and their product, the total weight [\equ{weight_total}], which is used to weight each event entering $\mathcal{P}_i(E_\nu)$.  Between different random mock flares, only the energy weights---and, by extension, the total weights---differ.

The mock flare simulation approach described above ensures that the subset of events with significant
directional weight, originally identified in~\cite{IceCube:2018cha} and re-evaluated in~\cite{IceCube:2021xar, IceCube:2023oua} as the most signal-like contributors to the 2014/2015 TXS 0506+056 neutrino flare, have the most influence on the irregularity, kurtosis, and skewness measures when we use ~\equ{rec1} to reconstruct the intrinsic emission timing pattern of the mock flare.

%%%%%%%%%%%%%%%%%%%%%%%%%%%%%%%%%%%%%%%%%%%%%%%%%%%%%%%%%%%%%%%%%%%%%%%%%%%%

\subsection{Statistical analysis}

%%%%%%%%%%%%%%%%%%%%%%%%%%%%%%%%%%%%%%%%%%%%%%%%%%%%%%%%%%%%%%%%%%%%%%%%%%%%
\begin{figure}[t!]
 \centering
 \includegraphics[width=1.0\columnwidth]{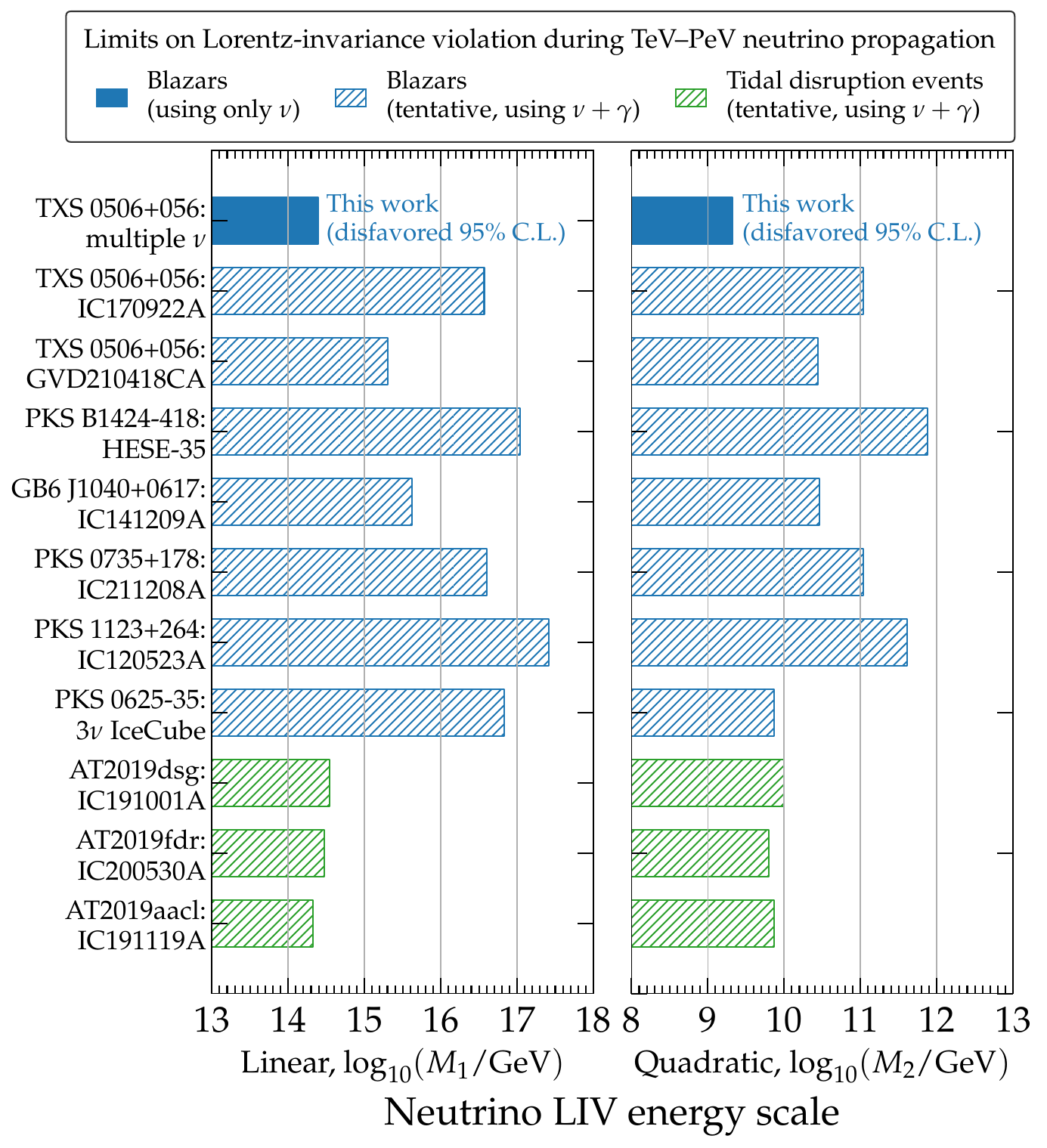}
 \caption{\label{fig:bounds_summary}
 \textbf{\textit{Limits on the energy scale of neutrino Lorentz-invariance violation.}} The results are derived from the observation of TeV--PeV neutrinos allegedly associated to a transient astrophysical emission episode, \ie, a flaring blazar or a tidal disruption event.
 \textbf{\textit{Our new results are robust upper limits derived solely from the time distribution of a flare of neutrinos detected by IceCube from the blazar TXS 0506+056 in 2014/2015.}}  We show the most conservative result from among Eqs.~(\ref{equ:limit_m1_ks})--(\ref{equ:limit_m1_skew}) for LIV linear in neutrino energy (\textit{left}) and from among Eqs.~(\ref{equ:limit_m2_ks})--(\ref{equ:limit_m2_skew}) for LIV quadratic in energy (\textit{right}).  See Secs.~\ref{sec:methods} and ~\ref{sec:liv_constraints_txs} for details.  All other results indicate tentative sensitivity, at an unspecified statistical significance, derived from the coincident detection of 1--3 neutrinos with electromagnetic emission from different candidate astrophysical sources.  They are marked as ``tentative'' because they hail from simpler analyses.  See Sec.~\ref{sec:other_limits} and Table~\ref{tab:IceCubenuLV} for details.}
\end{figure}
%%%%%%%%%%%%%%%%%%%%%%%%%%%%%%%%%%%%%%%%%%%%%%%%%%%%%%%%%%%%%%%%%%%%%%%%%%%%

The purpose of generating mock flares is to build from them the distribution of allowed values of the LIV measures, $\tau_n^{\rm KS}$, $\tau_n^{\rm kurt}$, and $\tau_n^{\rm skew}$ (Sec.~\ref{sec:methods-measures}).

For a given mock flare, we compute the LIV measures by maximizing the LIV estimators for irregularity ($\mathcal{D}$), kurtosis ($\mathcal{K}$), and skewness ($\mathcal{S}$), Eqs.~(\ref{equ:estimator_ks})--(\ref{equ:estimator_skew}).  In practice, we maximize the estimators by scanning $\tau_n$ over a fine grid of values that is also broad enough to avoid undesirable edge effects.  For each value of $\tau_n$, we compute the energy-dependent time lags, $b_{{\rm d}, i}$ in \equ{rec1}, and, with them, the estimators $\mathcal{D}$, $\mathcal{K}$, and $\mathcal{S}$.  Afterward, the positions on the grid where the estimators are maximized yield the LIV measures $\tau_n^{\rm KS}$, $\tau_n^{\rm kurt}$, and $\tau_n^{\rm skew}$.

We repeat this procedure many times, each time using a different random mock neutrino flare; in our results below, we use one million mock flares.  Doing this builds the probability distributions of the irregularity, kurtosis, and skewness measures, $\mathcal{P}(\tau_n^{\rm KS})$, $\mathcal{P}(\tau_n^{\rm kur})$, and $\mathcal{P}(\tau_n^{\rm skew})$, respectively.  These distributions allow us to assess whether the LIV measures are statistically compatible with zero values---allowing us to place limits on their values---or not---which would signal the presence of LIV.  We do this separately for LIV that is linear and quadratic in energy, \ie, for $n = 1$ and 2 (Sec.~\ref{sec:methods-delays}).

%%%%%%%%%%%%%%%%%%%%%%%%%%%%%%%%%%%%%%%%%%%%%%%%%%%%%%%%%%%%%%%%%%%%%%%%%%%%

\subsection{Results}
\label{sec:results}

Figure~\ref{fig:estimator_distributions} shows the resulting distributions of the LIV measures for $n = 1$ and 2.  In all cases, the distributions are compatible with $\tau_n^{\rm KS}$, $\tau_n^{\rm kurt}$, $\tau_n^{\rm skew}$ being zero.
Therefore, for each LIV measure, we place an upper limit on its value at the 95\% credible level (C.L.) by finding the maximum of its distribution and integrating the distribution around it. For example, in the case of the irregularity estimator, its allowed range is $\tau_n^{\rm KS} \leq \tau_{n,{\rm max}}^{\rm KS}$, where $\int_0^{\tau_{n,{\rm max}}^{\rm KS}} \mathcal{P}(\tau_n^{\rm KS}) = 0.95$, and similarly for the kurtosis and skewness measures. Figure~\ref{fig:estimator_distributions} shows the upper limits on the time-compensation parameters computed as above.

Figure~\ref{fig:bounds_summary} shows a summary of our main results, \ie, the lower limits (95\% C.L.) on the LIV energy scale, $M_1$ and $M_2$, derived from the upper limits on the LIV measures in \figu{estimator_distributions}.  We obtain them by inverting \equ{tauK1}:
\begin{eqnarray}
 \label{equ:limits_m1}
 && M_1^{\rm KS} \geq 2.5 \cdot 10^{14} ~\rm GeV \;, \label{equ:limit_m1_ks}\\
 && M_1^{\rm kur} \geq  3.9 \cdot 10^{14} ~\rm GeV\;, \label{equ:limit_m1_kur}\\
 && M_1^{\rm skew} \geq 4.0 \cdot 10^{15} ~\rm GeV \; \label{equ:limit_m1_skew},
\end{eqnarray}
for linear LIV, and
\begin{eqnarray}
 \label{equ:limits_m2}
 && M_2^{\rm KS} \geq 2.1 \cdot 10^{9} ~\rm GeV\;, \label{equ:limit_m2_ks}\\
 && M_2^{\rm kur} \geq  2.7 \cdot 10^{9} ~\rm GeV\;, \label{equ:limit_m2_kur}\\
 && M_2^{\rm skew} \geq 1.0 \cdot 10^{10} ~\rm GeV\;, \label{equ:limit_m2_skew}
\end{eqnarray}
for quadratic LIV.  To be conservative, \figu{bounds_summary} shows the smallest among them, separately for $n = 1$ and 2.

Taken at face value, \figu{bounds_summary} seemingly shows that our new limits are weaker than other limits also derived from the detection of transient high-energy neutrino emission.  This is, however, deceptive.  Other limits in \figu{bounds_summary} are marked ``tentative'' because, unlike ours, they rely on the coincident detection of neutrinos and electromagnetic emission---which, for many of the cases shown is speculative at best---and on simpler analyses that do not account for the significant uncertainty in the inferred energy of the detected neutrinos, nor for the angular separation between detected neutrinos and their alleged astrophysical source.  We elaborate on these shortcomings in Sec.~\ref{sec:other_limits}.

\textbf{\textit{Our new limits on the energy scale of Lorentz-invariance violation in neutrinos account for the experimental nuances associated to their detection and do not rely on the coincident detection of counterpart electromagnetic transient emission from the same astrophysical source.}}

%%%%%%%%%%%%%%%%%%%%%%%%%%%%%%%%%%%%%%%%%%%%%%%%%%%%%%%%%%%%%%%%%%%%%%%%%%%%
%%%%%%%%%%%%%%%%%%%%%%%%%%%%%%%%%%%%%%%%%%%%%%%%%%%%%%%%%%%%%%%%%%%%%%%%%%%%

\section{Other constraints on neutrino Lorentz-invariance violation}
\label{sec:other_limits}

We now discuss other searches for LIV effects in neutrinos, complementary to ours, including some that use high-energy astrophysical neutrinos.  We present a selection of them below, pointing out limitations in the latter that make them tentative in comparison to ours.  Reference~\cite{Kostelecky:2008ts} contains a comprehensive list of constraints on LIV derived from a variety of experiments.

%%%%%%%%%%%%%%%%%%%%%%%%%%%%%%%%%%%%%%%%%%%%%%%%%%%%%%%%%%%%%%%%%%%%%%%%%%%%

\subsection{Supernova and accelerator limits}
\label{sec:other_limits-sn_accel_nu}

Supernova neutrinos have energies of up to tens of MeV, while historically accelerator neutrinos have had energies of up to tens of GeV (newer experiments will reach TeV-scale energies~\cite{Feng:2022inv, FASER:2024ykc}).  Because these energies are lower than the TeV--PeV energies of the astrophysical neutrinos that we have used, they can only probe lower values of the LIV energy scale; see \equ{vg1}.

\smallskip

\textbf{\textit{Supernova SN1987A.---}}As discussed in \Refe~\cite{Ellis:2008fc}, the near-coincidences in the arrival times of the neutrinos
of different energies detected in the Kamiokande-II, IMB, and Baksan
experiments from the core collapse of SN1987A yield $M_{1} > 2.7 \cdot 10^{10} \; {\rm GeV}$ and $M_{2} > 4.6 \cdot 10^{4} \; {\rm GeV}$.  These limits were obtained assuming subluminal neutrino propagation~\cite{Amelino-Camelia:1997ieq, Cohen:2011hx}, like ours.

\smallskip

\textbf{\textit{Accelerator neutrinos.---}}A first study~\cite{Ellis:2008fc} of the velocities of accelerator neutrinos by the MINOS Collaboration~\cite{MINOS:2015iks} yielded $M_{1} > 10^{5} \; {\rm GeV}$ and $M_{2} > 600 \; {\rm GeV}$, assuming again subluminal neutrino propagation. Reference~\cite{MINOS:2012ozn} and references therein include searches for other possible signatures of LIV in neutrino propagation.

%%%%%%%%%%%%%%%%%%%%%%%%%%%%%%%%%%%%%%%%%%%%%%%%%%%%%%%%%%%%%%%%%%%%%%%%%%%%
\begingroup
\squeezetable
\begin{table*}[t!]
 \begin{ruledtabular}
  \caption{\label{tab:IceCubenuLV}\textbf{\textit{Constraints on and sensitivity to Lorentz-invariance violation in the propagation of high-energy astrophysical neutrinos.}} Results are on the energy scale of LIV [see \equ{vg1}]: $M_1$ for effects that are linearly dependent on neutrino energy and $M_2$ for effects that are quadratically dependent on it.  Figure~\ref{fig:bounds_summary} compares the limits graphically.  Our new results are lower limits on the energy scales.  We show here only the approximate range; see Eqs.~(\ref{equ:limit_m1_ks})--(\ref{equ:limit_m2_skew}) for exact results.  See Sec.~\ref{sec:liv_constraints_txs} for details.  All other results, shown in parentheses, represent tentative sensitivity derived from reports of multi-messenger neutrino-electromagnetic observations.  See Sec.~\ref{sec:other_limits} for details.}
  \vspace{0.2cm}
  \centering
  \renewcommand{\arraystretch}{1.3}
  \begin{tabular}{ccccccccccc}
   \multicolumn{2}{c}{Source} &
   \multirow{2}{*}{Redshift} &
   \multirow{2}{*}{Remark} &
   \multirow{2}{*}{\makecell{Neutrino\\telescope}} &
   \multirow{2}{*}{\makecell{Neutrino\\energy [TeV]}} &
   \multirow{2}{*}{\makecell{Time delay,\\$\Delta t$ [days]}} &
   \multirow{2}{*}{\makecell{$p$-value\\src.~assoc.}} &
   \multicolumn{3}{c}{Lower limit on LIV scale [GeV]}
   \\
   \cline{1-2}
   \cline{9-11}
   Type & Object & & & & & & &
   Linear, $M_{1}$ &
   Quadratic, $M_{2}$ &
   Ref.\footnote{\label{fnote5}These are the references where the constraints on Lorentz-invariance violation quoted here were derived.  For references on the claimed association between neutrinos and the sources, see Sec.~\ref{sec:other_limits}.  Blank entries (``$\cdots$'') represent the cases calculated in this paper from observations described in the text.} \\
   \hline
   AGN & TXS 0506+056 & 0.3365 & Multiple $\nu$\footnote{\label{fnote1}Main result of this paper, obtained from the 2014/2015 neutrino flare from TXS 0506+056, without an electromagnetic counterpart.} & IceCube & $\sim 10^2$--$10^3$ & $\sim 100$ & 0.8\% & $\sim 10^{15}$ & $\sim 10^{10}$ & This paper \\
   & TXS 0506+056 & 0.3365 & Single $\nu$ + $\gamma$\footnote{\label{fnote2}Tentative result only; see Sec.~\ref{sec:other_limits} for an explanation.} & IceCube & 290 & 10 & 0.3\% & $(3.7 \cdot 10^{16})$ & $(1.1 \cdot 10^{11})$ & \cite{Ellis:2018ogq} \\
   & TXS 0506+056 & 0.3365 & Single $\nu$ + radio\footref{fnote2} & Baikal-GVD & 220 & 200 & See text & ($\sim 2.0\cdot 10^{15}$) & ($ \sim 2.8\cdot 10^{10}$) & $\cdots$ \\
   & PKS B1424-418 & 1.522 & Single $\nu$ + $\gamma$\footref{fnote2} & IceCube & 2000 & 160 & 5\% & $(1.1 \cdot 10^{17})$ & $(7.6 \cdot 10^{11})$ & \cite{Wang:2016lne} \\
   & GB6~J1040+0617 & $\ge 0.7351$ & Single $\nu$ + $\gamma$\footref{fnote2} & IceCube & 100 &
   100 & 30\% & ($4.2 \cdot 10^{15}$) & ($2.9 \cdot 10^{10}$) & $\cdots$ \\
   & PKS 0735+178 & $\ge 0.424$ & $4 \nu$ + $\gamma$\footref{fnote2} & Multiple & $170$ & 10 & See text & $(4.0 \cdot 10^{16})$ & $(1.1 \cdot 10^{11})$ & $\cdots$ \\
   & PKS 1123+264 & 2.341 & Single $\nu$ + $\gamma$\footref{fnote2} & IceCube & $> 200$ & 10 & See text & $(2.6\cdot 10^{17})$ & $(4.1\cdot 10^{11})$ & $\cdots$ \\
   & {PKS 0625-35} & 0.055 & $3 \nu$ + radio\footref{fnote2} & IceCube & 63--302 & 1 & 3.6$\sigma$ & {$(6.8\cdot 10^{16})$} & {$(\sim 2.0\cdot 10^{11})$} & $\cdots$ \\
   \hline
   TDE & AT2019dsg & 0.051 & Single $\nu$ + $\gamma$\footref{fnote2} & IceCube & 200 & 150 & {TDE events} & $(3.5 \cdot 10^{14})$ & $(1.0 \cdot 10^{10})$ & $\cdots$ \\
   & AT2019fdr & 0.267 & Single $\nu$ + $\gamma$\footref{fnote2} & IceCube & 80 & 393 & {combined:} & $(3.0 \cdot 10^{14})$ & $(6.3 \cdot 10^{9})$ & $\cdots$ \\
   & AT2019aacl & 0.036 & Single $\nu$ + $\gamma$\footref{fnote2} & IceCube & 170 & 148 & $6 \cdot 10^{-4}$ & $(2.1 \cdot 10^{14})$ & $(7.4 \cdot 10^{9})$ & $\cdots$ \\
  \end{tabular} %\\
 \end{ruledtabular}
\end{table*}
\endgroup
%%%%%%%%%%%%%%%%%%%%%%%%%%%%%%%%%%%%%%%%%%%%%%%%%%%%%%%%%%%%%%%%%%%%%%%%%%%%

%%%%%%%%%%%%%%%%%%%%%%%%%%%%%%%%%%%%%%%%%%%%%%%%%%%%%%%%%%%%%%%%%%%%%%%%%%%%

\subsection{Challenges in associating neutrinos with electromagnetic signals for LIV constraints}
\label{sec:other_limits-challenges}

Limits on LIV effects during the propagation of high-energy neutrinos from identified astrophysical sources depend on the reliability of associating these neutrinos with their proposed sources. Establishing this association is challenging, which explains why only a few high-energy neutrino source candidates have been identified so far (though more will likely follow~\cite{Guepin:2022qpl}). This difficulty is particularly pronounced for transient astrophysical sources, which, due to their short duration, typically yield a very low number of detected neutrinos, often just a single one. Associating a single neutrino with a particular transient source thus relies on coincident observations of electromagnetic emissions from that source, often in gamma rays, as these are believed to be produced in the same particle interactions as neutrinos~\cite{Kelner:2006tc, Kelner:2008ke}.

When only a single neutrino is detected and is thought to originate from a specific source,
the derived limits on LIV must rely on the time delays between the detected neutrino and the
accompanying electromagnetic signal from the same source, rather than on the distribution of
neutrino arrival times, as used in our primary analysis
presented in Secs.~\ref{sec:methods} and~\ref{sec:liv_constraints_txs}.
This approach, however, introduces complications. Namely, a sizable time delay between the neutrino and electromagnetic
detection---a signal of potentially prominent LIV effects---weakens the association of the detected neutrino with the transient source, moreso since the neutrino and electromagnetic signals may have been emitted at substantially different times, without recourse to LIV.

To probe robustly LIV through the detection of a single neutrino in a transient multi-messenger emission episode, it is necessary to evaluate the association of said neutrino with the alleged source while taking into account possible differences in the emission times of high-energy neutrinos and photons, a challenging task that may require detailed source modeling~\cite{Levy:2024eiq}. Further, a comprehensive analysis should determine how large an LIV-induced time lag between the transient neutrino and electromagnetic emission could be while still maintaining a statistically meaningful association between them, especially in the presence of the background of atmospheric neutrinos and muons that would cloud the neutrino signal (Sec.~\ref{sec:methods-background}).

\textbf{\textit{LIV studies that fail to account for the above nuances could incorrectly attribute the observed time lag between transient neutrino and electromagnetic emission solely to LIV, and so claim unrealistic sensitivity to its effects.}}

Addressing the above complexities requires a dedicated analysis beyond the scope and goals of this paper.
Nonetheless, with these limitations in mind, we present below potential sensitivities to LIV derived
from claimed associations between single neutrinos (or a handful of them) and flaring blazars, TDEs, and GRBs.
In most cases, the associations of the neutrinos to these sources have low statistical significance, so that the identifications of
the multi-messenger emission episodes listed in Table~\ref{tab:IceCubenuLV} can only be considered as tentative, and hence also the derived sensitivity to LIV.

%%%%%%%%%%%%%%%%%%%%%%%%%%%%%%%%%%%%%%%%%%%%%%%%%%%%%%%%%%%%%%%%%%%%%%%%%%%%

\subsection{Tentative sensitivity from multi-messenger events: neutrinos and electromagnetic counterparts}
\label{sec:other_limits-nu_gamma_associations}

Table~\ref{tab:IceCubenuLV} summarizes the constraints on LIV based on claims of coincident detection of high-energy neutrinos and counterpart electromagnetic emission.
We expand on them below, keeping the above caveats in mind.  We indicate which of the limits we compute using Eqs.~(\ref{equ:delta_time}) and (\ref{equ:tauK1}), and the observed time delays and neutrino energies listed in Table~\ref{tab:IceCubenuLV}, and which were computed elsewhere~\footnote{In many of these analyses it is unclear how the neutrino energy used to compute the LIV effects was inferred from the detected muon energy proxy.} and are merely quoted here.  (The limits on LIV in neutrino propagation are generally weaker than in photon propagation~\cite{Kostelecky:2008ts}, but they are completely independent, as there is no general model-independent relation between LIV effects on neutrinos and photons.)

\smallskip

\textbf{\textit{Blazar TXS 0506+056.---}}In 2017, IceCube detected a single 290-TeV neutrino (IC170922A) from the direction of TXS 0506+056 in coincidence with an electromagnetic flare across multiple wavelengths, from radio to gamma rays~\cite{IceCube:2018dnn}.  In particular, about ten days after the neutrino detection, MAGIC detected gamma rays from the source~\cite{IceCube:2018dnn}.  Reference~\cite{Ellis:2018ogq} interpreted this difference in arrival times to find the limits $M_{1} > 3 \cdot 10^{16} \; {\rm GeV}$ and $M_{2} > 10^{11} \; {\rm GeV}$, assuming subluminal neutrino propagation.  Reference~\cite{Laha:2018hsh} found comparable limits via similar arguments. (The less precise temporal overlap of IC170922A with  {\it Fermi}-LAT data~\cite{Fermi-LAT:2019hte} would give a weaker constraint.)  The statistical significance of the association between the IceCube event and the gamma-ray flare of TXS 0506+056 is of about 3$\sigma$; most of the constraints drawn below from other astrophysical sources have lower statistical significance.\footnote{Reference~\cite{Wang:2020tej} constrained LIV in the case of superluminal propagation by demanding that the mean free path of this neutrino was not smaller than the distance to TXS 0506+056. However, this argument does not apply to subluminal propagation.}

\smallskip

\textbf{\textit{Radio flare from TXS 0506+056.---}}The Baikal-GVD Collaboration reported~\cite{Baikal-GVD:2022fmn} the detection of a high-energy neutrino (GVD210418CA) during a radio flare of TXS~0506+056 observed by RATAN-600, with a  duration of about three years.  Because the long duration of the radio flare weakens the association with the neutrino, the limits on LIV garnered from it are weaker than those coming from the 2017 TXS 0506+056 flare above~\cite{Ellis:2018ogq}.

\smallskip

\textbf{\textit{Blazar PKS B1424-418.---}}Reference ~\cite{Kadler:2016ygj} reported the association between an IceCube PeV-scale neutrino (HESE-35) and the gamma-ray blazar PKS B1424-418 during a 160-day flare.  However, the probability that this association could have arisen by chance is 5\%~\cite{Kadler:2016ygj}, high enough to render it questionable.  Regardless, based on this association, \Refe~\cite{Wang:2016lne} reported the limits on $M_1$ and $M_2$ that we quote in Table~\ref{tab:IceCubenuLV} and \figu{bounds_summary}.

\smallskip

\textbf{\textit{Blazar GB6~J1040+0617.---}}Reference~\cite{Fermi-LAT:2019hte} reported the association between an IceCube neutrino (IC141209A) and a flare of the blazar GB6~J1040+0617.  In this case, the probability of a chance coincidence is even higher, at 30\%.

\smallskip

\textbf{\textit{Blazar PKS~0735+178.---}}Reference~\cite{Sahakyan:2022nbz} reported the association between a multi-wavelength flare of the blazar PKS~0735+178 and four neutrino events detected by IceCube (170~TeV), Baikal (46~TeV), KM3NeT (16~TeV), and Baksan (4~GeV). The probability of a chance coincidence is approximately 3\%, and the position of PKS~0735+178 is outside the IceCube 90\% localization error, weakening the association.

\smallskip

\textbf{\textit{Blazar PKS 1123+264.---}}Reference~\cite{Hovatta:2020lor} reported a general study of possible coincidences between IceCube neutrinos and radio sources observed at the Owens Valley and Mets{\" a}hovi Radio Observatories, where it was found that observations of large radio flares from blazars in approximate coincidence with neutrino events are only due to chance. For our purpose, the most interesting coincidence is between a radio flare of PKS 1123+264 and an IceCube event (IC120523A) with an energy exceeding 200~TeV. The significance of this coincidence is weak, but not quantified in \Refe~\cite{Hovatta:2020lor}.

\smallskip

\textbf{\textit{Radio galaxy PKS 0625-35.---}}Reference~\cite{Bradascio:2023xha} reported the association of three IceCube neutrinos with energies between 63 and 302~TeV with a flare of the radio galaxy PKS 0625-35. The pre-trial significance of this apparent association was 3.6$\sigma$.

\smallskip

\textbf{\textit{Tidal disruption events.---}}Three IceCube high-energy neutrinos have been claimed to be coincident with electromagnetic emission from candidate TDEs.  One is a 200-TeV neutrino (IC191001A) observed 150 days after
the onset of emissions from TDE candidate AT2019dsg~\cite{Stein:2020xhk}, with a chance
coincidence probability of 0.5\%.  The second is an 80-TeV neutrino (IC200530A) observed 393 days
after the onset of emissions from TDE candidate AT2019fdr~\cite{Reusch:2021ztx}.  The third is a 170-TeV neutrino (IC191119A) observed 148 days after the onset of emissions from a TDE candidate AT2019aacl~\cite{vanVelzen:2021zsm,Zheng:2022kam}.
It is estimated~\cite{vanVelzen:2021zsm} that the significance of observing three chance neutrino-TDE coincidences is 3.6$\sigma$.  (In addition, two more IceCube neutrinos may be correlated with two obscured TDEs~\cite{Jiang:2023kbb}, with a chance coincidence of 3\%.)

\smallskip

\textbf{\textit{Gamma-ray bursts.---}}Reference~\cite{Amelino-Camelia:2022pja} reported a statistical analysis of the arrival times of IceCube neutrinos that may be associated with
gamma-ray bursts (GRBs), with an estimated sensitivity of $M_1 \approx 5 \cdot 10^{17}$~GeV (see also the previous analyses cited in \Refe~\cite{Amelino-Camelia:2022pja}).  The probability that the apparent delays in neutrino arrival times may be a ``false alarm'' is estimated to be 0.7\%.
Moreover, the redshifts of the GRBs were not measured, in general, and in the case of one IceCube
neutrino there were three possible candidates for the associated GRB.  Furthermore, all IceCube searches for neutrinos correlated with GRBs to date have yielded negative results, even when expanding the search window to 14~days around the prompt GRB emission~\cite{IceCube:2022rlk} and to lower energies~\cite{IceCube:2023woj}.  Therefore, we regard this result as
intriguing but as yet inconclusive, motivating future analyses employing a sample of unambiguous GRB candidates
with measured redshifts, and we do not include it in Table~\ref{tab:IceCubenuLV} or \figu{bounds_summary}.

%%%%%%%%%%%%%%%%%%%%%%%%%%%%%%%%%%%%%%%%%%%%%%%%%%%%%%%%%%%%%%%%%%%%%%%%%%%%%%%
%%%%%%%%%%%%%%%%%%%%%%%%%%%%%%%%%%%%%%%%%%%%%%%%%%%%%%%%%%%%%%%%%%%%%%%%%%%%%%%

\section{Summary}
\label{sec:summary}

The discovery of transient astrophysical emission of high-energy neutrinos with TeV--PeV energies has brought new insight not only into astrophysics, but also into fundamental physics, including making possible new, powerful tests of proposed extensions of the Standard Model.  We have shown how to use the observation of a high-energy astrophysical neutrino flare to test Lorentz invariance, a pillar of special and general relativity.

Because of Lorentz-invariance violation (LIV), neutrinos might travel slower than the speed of light, as suggested by a heuristic approach to quantum gravity~\cite{Amelino-Camelia:1997ieq}.  These effects are likely  suppressed by a high energy scale, but they are expected to grow linearly or quadratically with the energy of the neutrino and the distance it travels.  This makes high-energy astrophysical neutrinos---which travel Mpc--Gpc distances from their sources to Earth---potentially sensitive to LIV.

We have searched for LIV in the joint distribution of energies and arrival times of the neutrinos at Earth.  Its effect, if present, would be to render a burst of neutrinos more uniformly distributed in time upon reaching Earth, with higher-energy neutrinos affected more prominently.  Borrowing from searches of LIV in gamma rays, we use three statistical measures---of irregularity, kurtosis, and skewness of the neutrino time distribution---to quantify these effects and to place limits on the energy scale of LIV. Unlike previous searches that used high-energy astrophysical neutrinos detected in coincidence with an electromagnetic flare from the same astrophysical source, ours
relies only on the detection of the neutrino flare, without the need of an electromagnetic counterpart.

We illustrate our methods using the high-energy neutrino flare detected in 2014/2015 from the blazar TXS 0506+056 by the IceCube neutrino telescope.  These neutrinos were detected as through-going muon tracks, and there is appreciable uncertainty in inferring the energy of the neutrinos that created the muons, which we account for by generating a large number of simulated mock neutrino flares across the spread of possible neutrino energies.  In addition, we account indirectly for the presence of an irreducible background of atmospheric neutrinos and muons that could cloud the presence of LIV in the sample of neutrinos detected from a flare.

We find no evidence of LIV-induced distortions in the distribution of the arrival times of the neutrinos from TXS 0506+056, so we place new lower limits on the LIV energy scale; see
Eqs.~(\ref{equ:limit_m1_ks})--(\ref{equ:limit_m2_skew}) and \figu{bounds_summary}.
For LIV that depends linearly on neutrino energy, the limits (at 95\%~C.L.) are $M_1 \gtrsim 10^{14}$--$10^{15}$~GeV, depending on the measure used to derive them. For quadratically-dependent LIV, they are $M_2 \gtrsim 10^{9}$--$10^{10}$~GeV.  \textbf{\textit{Our new limits on the LIV energy scale account for realistic experimental nuance and do not rely on the coincident detection of neutrinos and electromagnetic counterparts.}}

Our limits are orders of magnitude stronger than those derived from accelerator neutrinos and observations of supernova SN1987A. For completeness, we have derived tentative sensitivities to LIV effects from several claims of multi-messenger association between single high-energy neutrinos and electromagnetic counterpart emission from an astrophysical source---a flaring blazar or a tidal disruption event---detected in coincidence. Some of these tentative sensitives appear superficially stronger than our limits. However, assessing whether these tentative sensitivities can be turned into robust limits requires dedicated study, and we emphasize that our limits are derived from a more robust statistical procedure.

As new neutrino telescopes around the world~\cite{Schumacher:2021hhm} start or ramp up operations~\cite{MammenAbraham:2022xoc, Ackermann:2022rqc, Guepin:2022qpl} we anticipate detecting more high-energy neutrino flares with which to test Lorentz invariance---and fundamental physics in general.

%%%%%%%%%%%%%%%%%%%%%%%%%%%%%%%%%%%%%%%%%%%%%%%%%%%%%%%%%%%%%%%%%%%%%%%%%%%%
%%%%%%%%%%%%%%%%%%%%%%%%%%%%%%%%%%%%%%%%%%%%%%%%%%%%%%%%%%%%%%%%%%%%%%%%%%%%

\medskip

\section*{Acknowledgements}

We thank Anna Franckowiak and Chad Finley for their help with the public IceCube data from TXS 0506+056.  A.S.S. expresses gratitude to Sergey Ganjour for his advice on RooFit. M.B. is supported by the {\sc Villum Fonden} under project no.~29388.
The work of J.E. was supported in part by STFC Grant ST/T000759/1.
The work of R.K. was partially supported by the Kakos Endowed Chair in Science Fellowship.  This work used the Tycho supercomputer hosted at the SCIENCE High Performance Computing Center at the University of Copenhagen.

%%%%%%%%%%%%%%%%%%%%%%%%%%%%%%%%%%%%%%%%%%%%%%%%%%%%%%%%%%%%%%%%%%%%%%%%%%%%
%%%%%%%%%%%%%%%%%%%%%%%%%%%%%%%%%%%%%%%%%%%%%%%%%%%%%%%%%%%%%%%%%%%%%%%%%%%%

%\bibliography{refs.bib}


%%%%%%%%%%%%%%%%%%%%%%%%%%%%%%%%%%%%%%%%%%%%%%%%%%%%%%%%%%%%%%%%%%%%%%%%%%%%
%%%%%%%%%%%%%%%%%%%%%%%%%%%%%%%%%%%%%%%%%%%%%%%%%%%%%%%%%%%%%%%%%%%%%%%%%%%%

\appendix

%%%%%%%%%%%%%%%%%%%%%%%%%%%%%%%%%%%%%%%%%%%%%%%%%%%%%%%%%%%%%%%%%%%%%%%%%%%%
%%%%%%%%%%%%%%%%%%%%%%%%%%%%%%%%%%%%%%%%%%%%%%%%%%%%%%%%%%%%%%%%%%%%%%%%%%%%

\section{Probability density function of the neutrino energy}
\label{sec:pdf_neutrino_energy}

\renewcommand{\theequation}{A\arabic{equation}}
\renewcommand{\thefigure}{A\arabic{figure}}
\renewcommand{\thetable}{A\arabic{table}}
\setcounter{figure}{0}
\setcounter{table}{0}

%%%%%%%%%%%%%%%%%%%%%%%%%%%%%%%%%%%%%%%%%%%%%%%%%%%%%%%%%%%%%%%%%%%%%%%%%%%%
\begin{figure*}[t!]
 \centering
 \includegraphics[width=\textwidth]{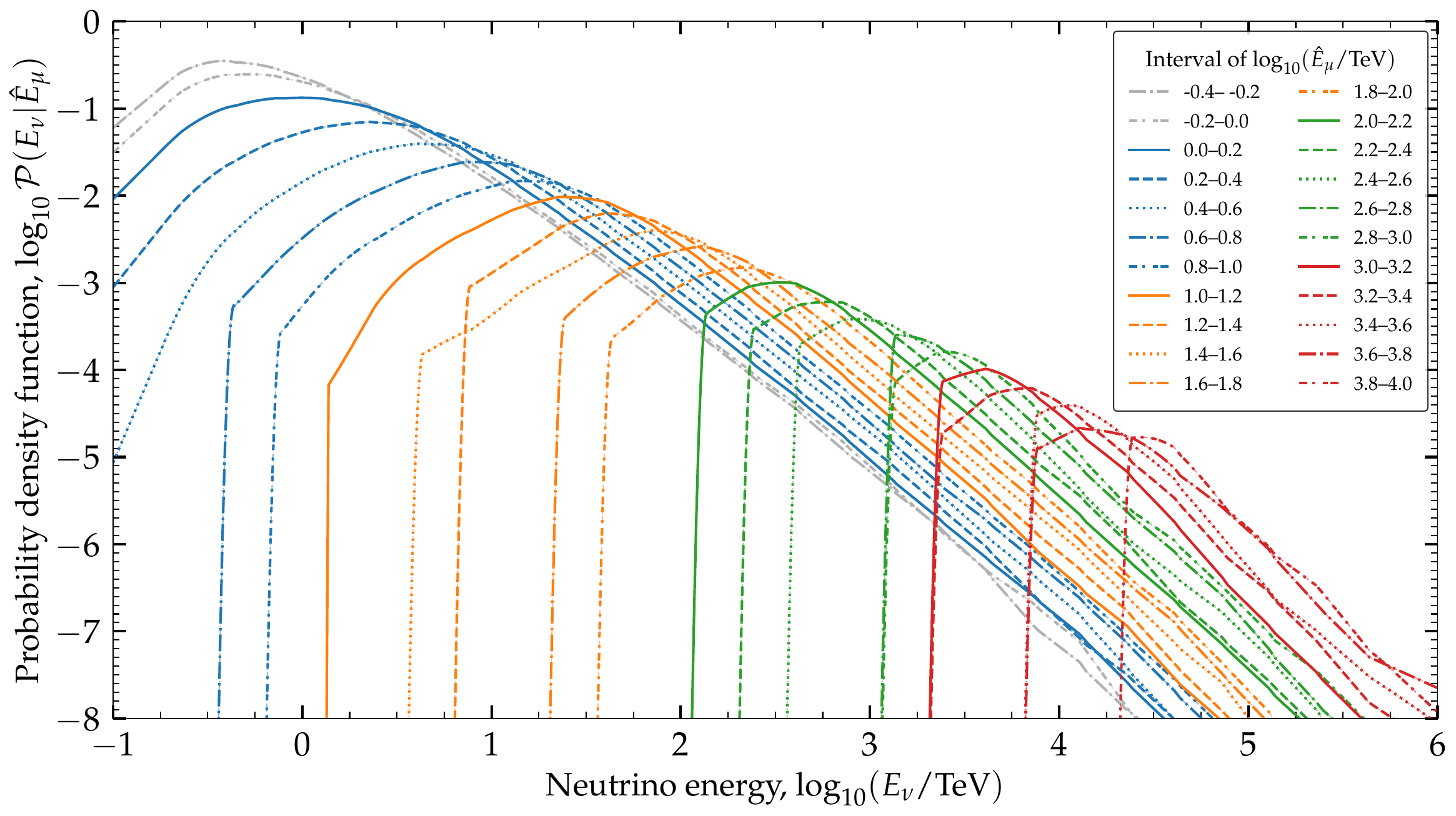}
 \caption{\label{fig:pdf_prob_energy_nu}\textbf{\textit{Probability density function, $\mathcal{P}(E_\nu \vert \hat{E}_\mu)$.}}  This is the first term in the integrand in \equ{pdf_energy_neutrino}, computed numerically using the data from \Refe~\cite{IceCube:2018cha} and then smoothed with a Savitzky-Golay filter~\cite{Golay:1964sg}.  See Appendix~\ref{sec:pdf_neutrino_energy} for details.}
\end{figure*}
%%%%%%%%%%%%%%%%%%%%%%%%%%%%%%%%%%%%%%%%%%%%%%%%%%%%%%%%%%%%%%%%%%%%%%%%%%%%

%%%%%%%%%%%%%%%%%%%%%%%%%%%%%%%%%%%%%%%%%%%%%%%%%%%%%%%%%%%%%%%%%%%%%%%%%%%%
\begin{figure*}[t!]
 \centering
 \includegraphics[width=0.5\textwidth]{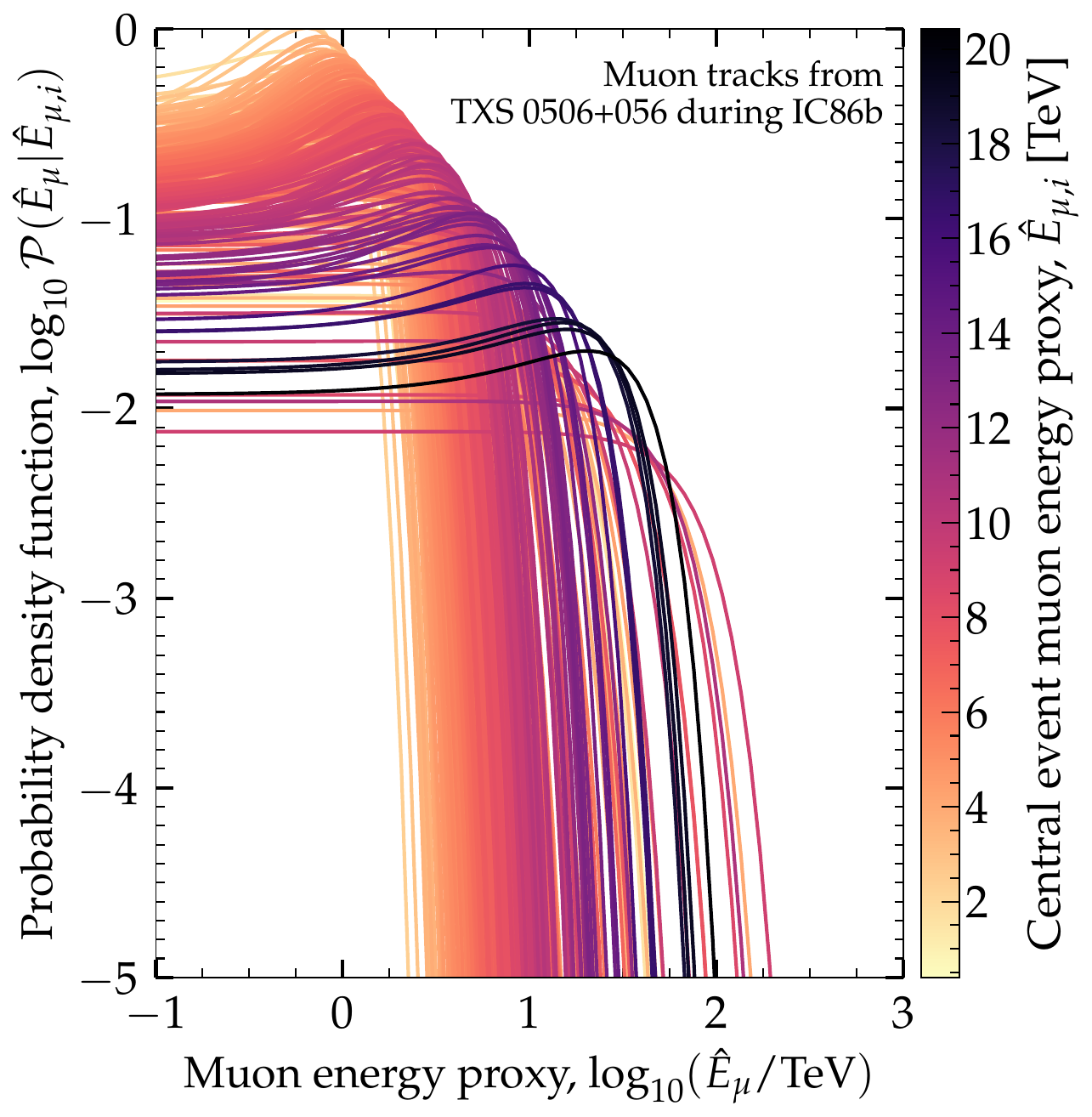}
 \caption{\label{fig:pdf_energy_proxy_ic86b}\textbf{\textit{Probability density function $\mathcal{P}(\hat{E}_\mu \vert \hat{E}_{\mu,i})$.}}  This is the second term in the integrand in \equ{pdf_energy_neutrino}, computed using \equ{pdf_muon_energy_proxy}, for all the muon tracks detected by IceCube from the direction of TXS 0506+056 during the IceCube observation period IC86b~\cite{IceCube:2018cha}.  We take the central values $\hat{E}_{\mu, i}$ from the IceCube public data release~\cite{IC_public_release}.  See Appendix~\ref{sec:pdf_neutrino_energy} for details.}
\end{figure*}
%%%%%%%%%%%%%%%%%%%%%%%%%%%%%%%%%%%%%%%%%%%%%%%%%%%%%%%%%%%%%%%%%%%%%%%%%%%%

Figure~\ref{fig:prob_energy_nu} in the main text shows the probability density function of the reconstructed neutrino energy, $\mathcal{P}_i(E_\nu)$ ($i = 1, \ldots, 320$), for each of the events detected by IceCube from the direction of TXS 0506+056 during the observation period IC86b~\cite{IceCube:2018cha}.  This is computed as
\begin{equation}
 \label{equ:pdf_energy_neutrino}
 \mathcal{P}_i(E_\nu)
 =
 \frac{1}{\mathcal{N}}
 \int_{\hat{E}_\mu^{\min}}^{\hat{E}_\mu^{\max}}
 \mathcal{P}(E_\nu \vert \hat{E}_\mu)
 \mathcal{P}(\hat{E}_\mu \vert \hat{E}_{\mu, i}) {\rm d} \hat{E}_\mu \; ,
\end{equation}
where $\hat{E}_{\mu, i}$ is the reported central value of the muon energy proxy of the $i$-th event.
Here, $\mathcal{N}$ is a normalization constant that ensures that $\int_0^\infty \mathcal{P}_i(E_\nu) {\rm d} E_\nu = 1$, $\mathcal{P}(E_\nu \vert \hat{E}_\mu)$ is the conditional probability distribution function that, given a measured muon energy proxy $\hat{E}_\mu$, the energy of the parent neutrino was $E_\nu$, and $\mathcal{P}(\hat{E}_\mu; \hat{E}_{\mu, i})$ is the probability distribution function of the muon energy proxy for this event. Below we expand on the two latter terms.

Figure~\ref{fig:pdf_prob_energy_nu} shows the first term in the integrand in \equ{pdf_energy_neutrino}, $\mathcal{P}(E_\nu \vert \hat{E}_\mu)$.  To compute it, we start from the data in Fig.~S4 of \Refe~\cite{IceCube:2018cha}, which shows $\mathcal{P}(\hat{E}_\mu \vert E_\nu )$.  (We use the tabulated data for Fig.~S4 in the IceCube public data release~\cite{IC_public_release} that accompanies \Refe~\cite{IceCube:2018cha}.)  We use these data to numerically build the cumulative distribution function $\mathcal{P}(E_\nu^\prime \leq E_\nu \vert \hat{E}_\mu)$.  From it, we compute the conditional probability distribution function that we seek, $\mathcal{P}(E_\nu \vert \hat{E}_\mu) = {\rm d}\mathcal{P}(E_\nu^\prime \leq E_\nu \vert \hat{E}_\mu)/{\rm d}E_\nu$.

Figure~\ref{fig:pdf_energy_proxy_ic86b} shows the second term in the integrand in \equ{pdf_energy_neutrino}, $\mathcal{P}(\hat{E}_\mu \vert \hat{E}_{\mu, i})$, computed for all of the muon tracks detected during the IC86b observation period~\cite{IceCube:2018cha}.
We model $\mathcal{P}(\hat{E}_\mu \vert \hat{E}_{\mu,i})$ as a normal distribution centered on the central value of the muon energy proxy of the event, $\hat{E}_{\mu,i}$, \ie,
\begin{equation}
 \label{equ:pdf_muon_energy_proxy}
 \mathcal{P}(\hat{E}_\mu \vert \hat{E}_{\mu,i})
 \equiv
 \frac{1}{\sqrt{2 \pi} \sigma(\hat{E}_{\mu,i})}
 \exp\left[
 -\frac{(\hat{E}_{\mu}-\hat{E}_{\mu,i})^2}{2 \sigma^2(\hat{E}_{\mu,i})}
 \right]
 \;,
\end{equation}
where we have taken the uncertainty in measuring the deposited energy to be 10\% in logarithmic scale~\cite{Aartsen:2013vja}, \ie,  $\sigma(\hat{E}_{\mu,i}) \equiv \max [ \lvert 10^{1.1 \log_{10}(\hat{E}_{\mu,i}/{\rm GeV})} - \hat{E}_{\mu,i} \rvert, \lvert \hat{E}_{\mu,i} - 10^{0.9 \log_{10}(\hat{E}_{\mu,i}/{\rm GeV})} \rvert ]$ in order to keep the largest between the lower and upper uncertainties.

%%%%%%%%%%%%%%%%%%%%%%%%%%%%%%%%%%%%%%%%%%%%%%%%%%%%%%%%%%%%%%%%%%%%%%%%%%%%%%%
%%%%%%%%%%%%%%%%%%%%%%%%%%%%%%%%%%%%%%%%%%%%%%%%%%%%%%%%%%%%%%%%%%%%%%%%%%%%%%%

\end{document}